\documentclass[article,tightenlines,showkeys,showpacs,amssymb,nofootinbib]{revtex4}
\usepackage{amsmath, amsthm, amscd, amssymb}
\usepackage{bm}
\usepackage{bbm}
\usepackage{graphicx}
\usepackage{color}
\usepackage{feyn}

\begin{document}

\title{Are Collapse Models Testable via Flavor Oscillations? %Collapse models and kaon oscillation
}
\author{Sandro Donadi}
\email{sandro.donadi@ts.infn.it}
\affiliation{Dipartimento di Fisica,
Universit\`a di Trieste, Strada Costiera 11, 34151 Trieste, Italy.
\\ Istituto Nazionale di Fisica Nucleare, Sezione di Trieste, Strada
Costiera 11, 34151 Trieste, Italy.}
\author{Angelo Bassi}
\email{bassi@ts.infn.it}
\affiliation{Dipartimento di Fisica,
Universit\`a di Trieste, Strada Costiera 11, 34151 Trieste, Italy.
 \\ Istituto Nazionale di Fisica Nucleare,
Sezione di Trieste, Strada Costiera 11, 34151 Trieste, Italy.}
\author{Catalina Curceanu}
\email{Catalina.Petrascu@lnf.infn.it}
\affiliation{Laboratori Nazionali di Frascati dell'INFN, Via E. Fermi 40, 00044 Frascati (Rome), Italy.}
\author{Antonio Di Domenico}
\email{antonio.didomenico@roma1.infn.it}
\affiliation{Dipartimento di Fisica, Sapienza Universit\`a di Roma, P.le Aldo Moro 5, 00185 Rome, Italy.
\\ Istituto Nazionale di Fisica Nucleare, Sezione di Roma, P.le Aldo Moro 5, 00185 Rome, Italy.}
\author{Beatrix C. Hiesmayr}
\email{beatrix.hiesmayr@univie.ac.at}
\affiliation{Masaryk University, Department of Theoretical Physics and Astrophysics, Kotl\'a\v{r}\'ska 2, 61137 Brno, Czech Republic.\\
University of Vienna, Faculty of Physics, Boltzmanngasse 5, 1090 Vienna, Austria.}

\begin{abstract}
Collapse models predict the spontaneous collapse of the wave function, in order to avoid the emergence of macroscopic superpositions. In their mass-dependent formulation,  they claim that the collapse of any system's wave function depends on its mass. Neutral K, D, B mesons are oscillating systems that are given by Nature as superposition of different mass eigenstates. Thus they are unique and interesting systems to look at, for analyzing the experimental implications of such models, so far in agreement with all known experiments. In this paper we derive---for the single mesons and bipartite entangled mesons---the effect of the mass-proportional CSL collapse model on the dynamics on neutral mesons, including the relativistic effects. We compare the theoretical prediction with experimental data from different accelerator facilities.
\end{abstract}
\pacs{03.65.-w, 03.65.Ud, 03.65.Ta}
\keywords{collapse models, particle-antiparticle oscillations}
 \maketitle
%\tableofcontents

\section{Introduction}

Flavored neutral mesons, those with a net non-zero strangeness, charm, or beauty content, are
among the most fascinating systems in elementary particle physics. Particle and antiparticle are distinguished only by the flavor quantum number, exhibiting the phenomenon of flavor oscillations in their time evolution
\cite{gm,pais} (see also~\cite{Beatrix1,Beatrix2,Bigi,Genovese,Durt,BenFlor,BenFlorRom,Beuthe}).
In the following we will focus on K-mesons, but our conclusions can be easily generalized to the other flavored meson systems, as discussed at the end of the paper.

Flavor oscillation is a direct consequence of the superposition principle of quantum mechanics. Nowadays many scientists question its validity~\cite{Bel,Ad,We,Le}, and several experiments have been performed~\cite{Ar1,Ar2,Ar3,Ar4,Bo,Mo} or proposed, which challenge it. On the theoretical side, models of spontaneous wave function collapse~\cite{Grw,Cslmass,Fu,Csl,adlerphoto,Pr,Im} explicitly predict that the superposition principle is valid only at the microscopic scale, while it gradually breaks down when moving towards the macroscopic scale. Therefore, by altering the standard quantum dynamics, collapse model predict a different behavior for kaon oscillations. Aim of this paper is to compute such an effect, and compare it with available experimental data. To this end, we will use one of the most popular collapse model, the mass-proportional CSL (Continuous Spontaneous Localization) model~\cite{Cslmass,Fu}, and will perform the calculation of the probabilities, that are observed in experiments, to second order perturbation theory and from that deduce the higher orders. Comparison with experimental data will allow to conclude whether current or future experiments are sensitive to this effect.

One of the difficulties in computing the effect of collapse models on kaon oscillation is that these models act on the {\it spatial} part of the wave function---since the collapse is supposed to localize wave functions in space---while oscillations occur for the internal degrees of freedom. What happens then, is that the noise responsible for the collapse of the wave function acts like a random medium through which the particles propagate, modifying their evolution and also the oscillatory behavior. By resorting to standard quantum field theoretical tools, we will compute analytically such an effect.

\section{Kaon Phenomenology and the mass-proportional CSL model}

We start by recalling why kaons oscillate. For convenience of the readers we keep the discussion here short and more details can be find in the Appendix~A and references therein. The basic observation is that the mass eigenstate are different from the flavour eigenstates. For kaons, there are two mass eigenstates, the short state $\left|K_{S}\right\rangle $ and the long state $\left|K_{L}\right\rangle $, as well as two flavor eigenstates, $\left|K^{0}\right\rangle $ and $\left|\bar{K}^{0}\right\rangle $. In this paper, we work in the approximation that there is no CP violation: this implies that $\left|K_{S}\right\rangle $ and $\left|K_{L}\right\rangle $ are orthogonal\footnote{This is justified by the smallness of the CP violation effect in neutral kaons, which gives
rise to a small (of the order of $10^{-3}$) odd/even CP impurity in the $K_S/K_L$ states, and
to a small non-orthogonality between them.}. The relation between
the mass and the flavour eigenstates is given by:
\begin{equation}
\left|K^{0}\right\rangle =\frac{\left|K_{L}\right\rangle +\left|K_{S}\right\rangle }{\sqrt{2}},\qquad\qquad\;\;\;\left|\bar{K}^{0}\right\rangle =\frac{\left|K_{L}\right\rangle -\left|K_{S}\right\rangle }{\sqrt{2}}.
\end{equation}
During the time evolution, the mass eigenstates change by acquiring different phase factors, depending on their mass\footnote{Indeed, this is not exactly true: if we take into account that kaons decay in time, there is also an exponential dumping factor, besides the phase factor. We will consider this property later. Here we want to explain the basic idea behind the oscillation phenomenon.}. The phenomenon of flavor oscillation arises because what we measure are not the mass eigenstates, but the flavor eigenstates, which are superpositions of mass eigenstates. The different phase factors in front of the mass eigenstates change the superposition, making it possible, for example, to start with a $\left|K^{0}\right\rangle$ and end up with a $\left|\bar{K}^{0}\right\rangle$.

According to collapse models, the time evolution of the mass eigenstates is different compared to the one given by standard quantum mechanics. Hence we expect to see a different behavior in kaon oscillations. According to the mass proportional CSL model~\cite{Cslmass,Fu}, the evolution of the state vector is given by the non-linear equation:
\begin{equation} \label{eq:csl-massa}
d|\phi_{t}\rangle=\left[-\frac{i}{\hbar}Hdt+\frac{\sqrt{\gamma}}{m_{0}}\int d\mathbf{x}\,\left(M(\mathbf{x})-\left\langle M(\mathbf{x})\right\rangle \right)dW_{t}(\mathbf{x})-\frac{\gamma}{2m_{0}^{2}}\int d\mathbf{x}\,\left(M(\mathbf{x})-\left\langle M(\mathbf{x})\right\rangle \right)^{2}dt\right]|\phi_{t}\rangle\;,
\end{equation}
where $\left\langle M(\mathbf{x})\right\rangle :=\left\langle \phi_{t}\left|M(\mathbf{x})\right|\phi_{t}\right\rangle $. Here $H$ is the standard quantum Hamiltonian of the system and the other two
terms induce the collapse of the wave function in space. The mass $m_0$ is a reference mass, which is taken
equal to that of a nucleon. The parameter $\gamma$ is a positive coupling
constant which sets the strength of the collapse process, while $M({\bf x})$ is
a smeared mass density operator:
\begin{equation}
M\left(\mathbf{x}\right)=\underset{j}{\sum}m_{j}N_{j}\left(\mathbf{x}\right)\;,\qquad
N_{j}\left(\mathbf{x}\right)=\int
d\mathbf{y}g\left(\mathbf{y-x}\right)
\psi_{j}^{\dagger}\left(\mathbf{y}\right)\psi_{j}\left(\mathbf{y}\right)\;,
\end{equation}
where $\psi_{j}^{\dagger}\left(\mathbf{y}\right)$,
$\psi_{j}\left(\mathbf{y}\right)$ are, respectively, the creation and
annihilation operators of a particle of type $j$, namely having mass $m_j$ and spin $s$, in the space point $\mathbf{y}$. Neutral kaons are spin zero particles, thus the spin will be of no relevance for the following calculations. The smearing function $g({\bf x})$ is usually taken to be a Gaussian:
\begin{equation} \label{eq:nnbnm}
g(\mathbf{x}) \; = \; \frac{1}{\left(\sqrt{2\pi}r_{C}\right)^{3}}\;
e^{-\mathbf{x}^{2}/2r_{C}^{2}}\;,
\end{equation}
where $r_C$ is the second new phenomenological constant of the model. The standard numerical value of this correlation length $r_C$ is~\cite{Grw,Csl,Pr}:
\begin{equation}
r_C \; = \; 10^{-5}\text{cm},
\end{equation}
while, in the literature, two different values for the collapse strength $\gamma$ have been proposed. The first value has been originally proposed by Ghirardi, Pearle and Rimini~\cite{Csl}
\begin{equation}
\gamma \; = \; 10^{-30}\text{cm}^{3}\text{s}^{-1}
\end{equation}
in analogy with the GRW model~\cite{Grw}. The second value has been proposed by Adler, inspired by the analysis of the process of latent image formation according to collapse models, and amounts to~\cite{adlerphoto}
\begin{equation}
\gamma \; = \; 10^{-22}\text{cm}^{3}\text{s}^{-1}.
\end{equation}
Finally, $W_{t}\left(\mathbf{x}\right)$ is an
ensemble of independent Wiener processes, one for each point in space.

Working with non-linear equations is notoriously difficult. As shown e.g. in Ref.~\cite{Im} the experimentally testable predictions of the model---when averaged over the noise---do not change if the real noise $W_{t}\left(\mathbf{x}\right)$ is replaced by an imaginary noise
$i W_{t}\left(\mathbf{x}\right)$. In this way, one loses the collapse properties, i.e. the non-linear terms, of the equation. However, the advantage of having an imaginary noise is that the evolution is described by a standard Schr\"odinger equation with a random Hamiltonian:
\begin{equation} \label{eq:htot}
H_{\text{\tiny TOT}} = H - \hbar \sqrt{\gamma} \sum_j \frac{m_{j}}{m_{0}}
\int d\mathbf{y}\, w(\mathbf{y},t)
\psi_{j}^{\dagger}(\mathbf{y}) \psi_{j}(\mathbf{y})
\end{equation}
where
\begin{equation} \label{eq:sfsoi}
w(\mathbf{y},t) = \int d\mathbf{x}\, g(\mathbf{y-x})\xi_{t}(\mathbf{x}),
\end{equation}
and $\xi_{t}(\mathbf{x}) = dW_{t}(\mathbf{x})/dt$ is a white noise field, with
correlation function $\mathbb{E}[ \xi_{t}(\mathbf{x}) \xi_{s}(\mathbf{y})] =
\delta(t-s) \delta({\bf x-y})$. As such, $w(\mathbf{x},t)$ is a Gaussian noise
field, with zero mean and the correlation function:
\begin{equation} \label{eq:sdfddas}
{\mathbb E}[w(\mathbf{x},t) w(\mathbf{y},s)] \; = \;
\delta(t-s)F({\bf x} - {\bf y}), \qquad  F({\bf x}) \; = \;
\frac{1}{(\sqrt{4 \pi} r_C)^3} e^{-{\bf x}^2/4 r_C^2}\;.
\end{equation}
In the following, we will be a bit more general without making the calculation more complicated, and we will assume that the noise has a general time correlation function $f(s)$ instead of a white noise time-correlator $\delta(s)$.

In our analysis, we treat the kaons as non-relativistic particles, which is in accordance with the experimental situation at some acceleration facilities. Accordingly, the Hamiltonian for the mass eigenstates is given by ($j=S,L$; short or long)
\begin{equation}
H\left(t\right) \; = \; \sum_{j=S,L}\int d\mathbf{x}\,\mathcal{H}_{j}\left(x\right)
\end{equation}
with
\begin{equation}
\mathcal{H}_{j}\left(x\right)={\underbrace{\displaystyle m_{j}c^{2}\psi_{j}^{\dagger}\left(x\right)\psi_{j}\left(x\right)+
\frac{\hbar^{2}}{2m_{j}}\nabla\psi_{j}^{\dagger}\left(x\right)\cdot\nabla\psi_{j}\left(x\right)}_{\displaystyle :=\mathcal{H}_{\mathcal{S}}^j\left(x\right)}}
-\underbrace{\displaystyle \hbar\sqrt{\gamma_{m_{j}}}w\left(x\right)\psi_{j}^{\dagger}\left(x\right)\psi_{j}\left(x\right)}_{\displaystyle  :=\mathcal{N}^{j}\left(x\right)}
\end{equation}
and
\begin{equation}\label{eq:gamma}
\gamma_{m_{j}}\equiv\gamma\left(\frac{m_{j}}{m_{0}}\right)^{2}\;.
\end{equation}
The Hamiltonian
includes two contributions: the standard Schr\"odinger term $\mathcal{H}_{\mathcal{S}}\left(x\right)=\mathcal{H}_{\mathcal{S}}^S\left(x\right)+\mathcal{H}_{\mathcal{S}}^L\left(x\right)$
and the term $\mathcal{N}\left(x\right)=\mathcal{N}^{S}\left(x\right)+\mathcal{N}^{L}\left(x\right)$ which accounts for the
collapse. Note that in $\mathcal{H}_{S}\left(x\right)$ we have included
also the mass energy terms. These terms are usually ignored, since they lead only to a constant shift of the energy, which implies no observable effect. In the specific situation of kaons, where we have a superposition of two different mass eigenstates, it is fundamental to keep it.

Indeed, for most investigations of the kaon phenomenology the kinetic part is not relevant and one introduces an effective Hamiltonian with two different mass eigenstates, i.e. the system is treated as a two state system analogously to spin-$\frac{1}{2}$ particles~\cite{Beatrix1,Beatrix2} and the two relevant bases---if $\mathcal{CP}$ violation is neglected---are the strangeness (flavor) basis and mass eigenstate basis. For convenience of the readers we provide a summary in the Appendix~A.

Since the effect of the CSL noise is to localize the wavefunction in space, we cannot simplify our model by neglecting the kinetic part and we need to work in a standard field theoretical background. In order to compute the effect of collapse models on their evolution, we must specify therefore the spatial part of the wave function, the form of the wave function in space, which will be localized via the background field. This means that, as said before, the change in the oscillatory behavior of kaons, according to the collapse dynamics, is an indirect, not a direct, effect of the collapse process.

In the following, we focus our attention on computing the probability that, starting
from a $\left|K^{0}\right\rangle $ state at time $t=0$, we end up in a $\left|\bar{K}^{0}\right\rangle$ state at a later time.
The others possible transition probabilities ($\left|K^{0}\right\rangle \longrightarrow\left|K^{0}\right\rangle $, $\left|\bar{K}^{0}\right\rangle \longrightarrow\left|\bar{K}^{0}\right\rangle $
and $\left|\bar{K}^{0}\right\rangle \longrightarrow\left|K^{0}\right\rangle $)
can be computed in a very similar way. In order to keep the computation as simple as possible, we assume that the initial state is a plane wave with definite momentum $\mathbf{p}_{i}$. Therefore, the quantity we wish to compute is:
\begin{equation}\label{eq:firstPk0}
P_{K^{0}\rightarrow\bar{K}^{0}}\left(\mathbf{p}_{i}\right) = \sum_{\mathbf{p}_{f}}\mathbb{E}\left|\left\langle \bar{K}^{0},\mathbf{p}_{f}\left|U\left(t\right)\right|K^{0},\mathbf{p}_{i}\right\rangle \right|^{2}
\end{equation}
where we assumed that also the final state is a momentum eigenstate, and we sum over all the possible final states. Here, $\mathbb{E}$ denotes the stochastic average with respect to the noise of the background field. It is convenient to express the matrix elements in terms of the mass eigenstates, instead of the strangeness eigenstates, since they are the diagonal states of the Hamiltonian and, therefore, provide the simplest form of the time evolution:
\begin{eqnarray}\label{eq:matelemk0}
\left\langle \bar{K}^{0},\mathbf{p}_{f}\left|U\left(t\right)\right|K^{0},\mathbf{p}_{i}\right\rangle & = & \sum_{i,j}\alpha_{j}\beta_{i}^{*}\left\langle K_{i},\mathbf{p}_{f}\left|U\left(t\right)\right|K_{j},\mathbf{p}_{i}\right\rangle =\sum_{j}\alpha_{j}\beta_{j}^{*}\left\langle K_{j},\mathbf{p}_{f}\left|U_{j}\left(t\right)\right|K_{j},\mathbf{p}_{i}\right\rangle \nonumber \\
& \equiv & \sum_{j}\alpha_{j}\beta_{j}^{*}T_{j}\left(\mathbf{p}_{f},\mathbf{p}_{i},t\right)\;,
\end{eqnarray}
where $i,j=S,L$ and $\alpha_{S}=\alpha_{L}=\beta_{L}=1/\sqrt{2}$, $\beta_{S}=-1/\sqrt{2}$ [The parameters $\alpha, \beta$ change accordingly for the other transition probabilities]. The second equation is due to the fact that since the structure of the Hamiltonian is given by $H\left(t\right)=H_{S}\left(t\right)+H_{L}\left(t\right)$, namely not mixing of the mass eigenstates, the time evolution operator factorizes $U\left(t\right)=U_{S}\left(t\right)\otimes U_{L}\left(t\right)$. This implies that if a mass eigenstate is produced at time $t=0$ it persists its identity for any later time point. This is strongly supported by experimental data.

%% A crucial property which considerably simplifies the calculation and that we used in going from the second to the third expression of Eq.~\eqref{eq:matelemk0}, is that the evolution operator $U\left(t\right)=U_{S}\left(t\right)\otimes U_{L}\left(t\right)$ is factorized, due to the structure of the Hamiltonian $H\left(t\right)=H_{S}\left(t\right)+H_{L}\left(t\right)$, which does not contain mixed terms involving $S$ and $L$ states. This implies that a {}``$j$'' state (mass eigenstate with mass $m_{j}$) can evolve only in the same ``$j$'' state.

Substituting Eq.~\eqref{eq:matelemk0} into Eq.~\eqref{eq:firstPk0} we get
\begin{equation} \label{eq:sdgsetj}
P_{K^{0}\rightarrow\bar{K}^{0}}\left(\mathbf{p}_{i},t\right)=\sum_{j,k}\alpha_{j}\beta_{j}^{*}\alpha_{k}^{*}\beta_{k}\underset{\displaystyle\equiv P_{kj}\left(\mathbf{p}_{i};t\right)}{\underbrace{\sum_{\mathbf{p}_{f}}\mathbb{E}\left[T_{k}^{*}\left(\mathbf{p}_{f},\mathbf{p}_{i},t\right)T_{j}\left(\mathbf{p}_{f},\mathbf{p}_{i},t\right)\right]}}.
\end{equation}
We now compute these probabilities.

\section{Derivation of the Probabilities for One-particle States}

We cannot solve the equations of motion exactly, because of the noise term in the Hamiltonian. Therefore, we move to the interaction picture and apply a standard perturbative approach. We treat $\mathcal{H}_{\mathcal{S}}\left(x\right)$ as the unperturbed Hamiltonian, and $\mathcal{N}\left(x\right)$ as a perturbation. This is certainly a very reasonable assumption since the noise coupling constant $\gamma$ is very small.

\subsection{The Interaction Picture}

In the interaction picture, the states evolve as follows~\cite{Sakurai}
\begin{equation}
\left|\psi_{t}\right\rangle _{I}\equiv U_{I}\left(t,0\right)\left|\psi_{0}\right\rangle _{I}\;,
\end{equation}
with
\begin{equation} \label{eq:dgr}
U_{I}\left(t,0\right)=1+\sum_{n=1}^{\infty}\left(\frac{-i}{\hbar}\right)^{n}\int_{0}^{t}dt_{1}\int_{0}^{t_{1}}dt_{2}...\int_{0}^{t_{n-1}}dt_{n}N_{I}\left(t_{1}\right)N_{I}\left(t_{2}\right)...N_{I}\left(t_{n}\right)
\end{equation}
given via the well known Dyson series. The relation between the evolution operator in the Schr\"odinger picture and the corresponding one in the interaction picture is given by
\begin{equation}
U\left(t,0\right)=e^{-\frac{i}{\hbar}H_{S}t}U_{I}\left(t,0\right).
\end{equation}
The fields, on the other hand, evolve accordingly to the free Hamiltonian, therefore we expand the wave function into a superposition of plane waves
\begin{equation} \label{eq:dgdy}
\psi_{I}\left(x\right)=\frac{1}{\sqrt{L^{3}}}\sum_{\mathbf{k}}b_{\mathbf{k}}e^{-\frac{i}{\hbar}\left(E_{k}t-\mathbf{k}\cdot\mathbf{x}\right)}\;.
\end{equation}
In order to avoid possible divergences coming from the fact that we are working with plane waves, we quantize our system in a box with side $L$ by imposing periodic boundary conditions. This implies that the components of the wave vector $\bf k$ take only discrete values $k_{j}=\frac{2\pi\hbar}{L}n_{kj}$, with $n_{kj}$ integer. The energy is $E_{k}=mc^{2}+\frac{\mathbf{k}^{2}}{2m}$ and $b_{\mathbf{k}}$ is the annihilation operator of a particle with momentum $\bf k$. Since kaons are bosons, their operators have to satisfy the commutation relations $[b_{\mathbf{k}},b_{\mathbf{k}'}^{\dagger}]=\delta_{\mathbf{k},\mathbf{k}'}$. The total Hamiltonian can then be written as \begin{equation}H = \sum_{\mathbf{k}}E_{k}\;b_{\mathbf{k}}^{\dagger}b_{\mathbf{k}}\;.\end{equation}

At the end of the calculation, we take the limit $L\rightarrow\infty$, which amounts to make the substitutions:
\begin{equation}
\sum_{\mathbf{k}=-\infty}^{+\infty}\longrightarrow\int d\mathbf{k}, \qquad\qquad \frac{1}{L^{3}}\longrightarrow\frac{1}{\left(2\pi\right)^{3}}.
\end{equation}
We have now introduced all the necessary elements for computing the matrix elements and the transition probabilities.

\subsection{Computation of the Transition Amplitudes}

We start by focusing our attention on the computation of the transition amplitude of a certain mass eigenstate $T_{j}\left(\mathbf{p}_{f},\mathbf{p}_{i},t\right)$. Since this computation is independent of the mass eigenstate $j=S,L$, we drop this index in this section. The interesting expression to compute is
\begin{equation}
T\left(\mathbf{p}_{f},\mathbf{p}_{i},t\right) \equiv \langle \mathbf{p}_{f} |U\left(t\right) |\mathbf{p}_{i}\rangle = \langle \mathbf{p}_{f} |e^{-\frac{i}{\hbar}H_{S}t}U_{I}\left(t,0\right) |\mathbf{p}_{i}\rangle = e^{-\frac{i}{\hbar}E_{f}t} \langle \Omega |b_{\mathbf{p}_{f}}U_{I}\left(t,0\right)b_{\mathbf{p}_{i}}^{\dagger} |\Omega \rangle,
\end{equation}
where $E_{f}=\frac{\mathbf{p}_{f}^{2}}{2m}$ and $|\Omega\rangle $ is the vacuum state. The perturbative scheme goes as follows. We keep terms up to second order, which---when averaged over the noise---give the first order contribution to the oscillation. The reason is that the corrections are proportional to the average of products of the noise computed in different space time points, which are not zero only when the noise appears at least twice in the matrix elements. Therefore we have
\begin{equation}
U_{I}\left(t,0\right)\simeq1-\frac{i}{\hbar}\int_{0}^{t}dt_{1}N_{I}\left(t_{1}\right)-\frac{1}{\hbar^{2}}\int_{0}^{t}dt_{1}\int_{0}^{t_{1}}dt_{2}N_{I}\left(t_{1}\right)N_{I}\left(t_{2}\right)
\;.
\end{equation}
Accordingly, the transition probability becomes:
\begin{equation}
\label{eq:dfgfdgfwe}
T\left(\mathbf{p}_{f},\mathbf{p}_{i},t\right)\simeq e^{-\frac{i}{\hbar}E_{f}t}\left[T^{\left(0\right)}\left(\mathbf{p}_{f};\mathbf{p}_{i};t\right)+T^{\left(1\right)}\left(\mathbf{p}_{f};\mathbf{p}_{i};t\right)+T^{\left(2\right)}\left(\mathbf{p}_{f};\mathbf{p}_{i};t\right)\right],
\end{equation}
where each term corresponds to one of the first three terms of the Dyson series. It is possible to give a representation by means of Feynman diagrams for each one of these three terms:
\[
\begin{array}{cclcclccl}
\Diagram{\vertexlabel^{i}fsfAfs\vertexlabel^{f}} & = & \displaystyle  T^{\left(0\right)}\left(\mathbf{p}_{f};\mathbf{p}_{i};t\right), \qquad\qquad &
\Diagram{\vertexlabel^ifA hu \\ f0 fdA\vertexlabel^{\;\;f}} & = & \displaystyle T^{\left(1\right)}\left(\mathbf{p}_{f};\mathbf{p}_{i};t\right), \qquad\qquad &
\Diagram{\vertexlabel_i hd\vertexlabel_{1}  fA \vertexlabel_{2}
hu\\
fuA f0 fdA \vertexlabel^{\;\;f}}  & = & \displaystyle T^{\left(2\right)}\left(\mathbf{p}_{f};\mathbf{p}_{i};t\right),
\end{array}
\]
where the solid lines refer to the kaon and the dotted lines to the noise.
The first term, $T^{\left(0\right)}$, is trivial:
\begin{equation}
T^{\left(0\right)}\left(\mathbf{p}_{f};\mathbf{p}_{i};t\right) \equiv \langle \Omega |b_{\mathbf{p}_{f}}b_{\mathbf{p}_{i}}^{\dagger} |\Omega \rangle = \delta_{\mathbf{p}_{f},\mathbf{p}_{i}}.
\end{equation}
The second term is given by:
\begin{equation}\label{eq:T1}
T^{\left(1\right)}\left(\mathbf{p}_{f};\mathbf{p}_{i};t\right) \equiv i\sqrt{\gamma_{m}}\int_{0}^{t}dt_{1}\int d\mathbf{x}_{1}w\left(x_{1}\right) \langle \Omega |b_{\mathbf{p}_{f}}\psi_{I}^{\dagger}\left(x_{1}\right)\psi_{I}\left(x_{1}\right)b_{\mathbf{p}_{i}}^{\dagger} |\Omega \rangle.
\end{equation}
Using the plane-wave expansion of the fields given by Eq.~\eqref{eq:dgdy}, and a similar expansion for the adjoint of the field, we get:
\begin{eqnarray}
\langle \Omega |b_{\mathbf{p}_{f}}\psi_{I}^{\dagger}\left(x_{1}\right)\psi_{I}\left(x_{1}\right)b_{\mathbf{p}_{i}}^{\dagger} |\Omega \rangle & = &
\frac{1}{L^{3}}\sum_{\mathbf{k},\mathbf{k}'}e^{\frac{i}{\hbar}\left(E_{k'}t_{1}-\mathbf{k}'\cdot\mathbf{x}_{1}\right)}e^{-\frac{i}{\hbar}\left(E_{k}t_{1}-\mathbf{k}\cdot\mathbf{x}_{1}\right)} \langle \Omega |b_{\mathbf{p}_{f}}b_{\mathbf{k}'}^{\dagger}b_{\mathbf{k}}b_{\mathbf{p}_{i}}^{\dagger} |\Omega \rangle \nonumber \\
& = & \frac{1}{L^{3}}\;e^{\frac{i}{\hbar}\left[\left(E_{f}-E_{i}\right)t_{1}-\left(\mathbf{p}_{f}-\mathbf{p}_{i}\right)\cdot\mathbf{x}_{1}\right]},
\end{eqnarray}
and therefore:
\begin{equation}
T^{\left(1\right)}\left(\mathbf{p}_{f};\mathbf{p}_{i};t\right)=i\sqrt{\gamma_{m}}\frac{1}{L^{3}}\int_{0}^{t}dt_{1}\int d\mathbf{x}_{1}w\left(x_{1}\right)e^{\frac{i}{\hbar}\left[\left(E_{f}-E_{i}\right)t_{1}-\left(\mathbf{p}_{f}-\mathbf{p}_{i}\right)\cdot\mathbf{x}_{1}\right]}.
\end{equation}
The third term is given by:
\begin{eqnarray}
T^{\left(2\right)}\left(\mathbf{p}_{f};\mathbf{p}_{i};t\right) & = & -\gamma_{m}\int_{0}^{t}dt_{1}\int_{0}^{t_{1}}dt_{2}\int d\mathbf{x}_{1}\int d\mathbf{x}_{2}w\left(x_{1}\right)w\left(x_{2}\right) \nonumber \\
& & \cdot  \langle \Omega |b_{\mathbf{p}_{f}}\psi_{I}^{\dagger}\left(x_{1}\right)\psi_{I}\left(x_{1}\right)\psi_{I}^{\dagger}\left(x_{2}\right)\psi_{I}\left(x_{2}\right)b_{\mathbf{p}_{i}}^{\dagger} |\Omega \rangle\;.
\end{eqnarray}
The matrix element becomes:
\begin{eqnarray}
\lefteqn{\langle \Omega |b_{\mathbf{p}_{f}}\psi_{I}^{\dagger}\left(x_{1}\right)\psi_{I}\left(x_{1}\right)\psi_{I}^{\dagger}\left(x_{2}\right)\psi_{I}\left(x_{2}\right)b_{\mathbf{p}_{i}}^{\dagger} |\Omega \rangle =} \qquad\qquad\qquad\nonumber \\
& = & \frac{1}{L^{3}}\;\sum_{\mathbf{k},\mathbf{k}'}e^{-\frac{i}{\hbar}\left(E_{k}t_{2}-\mathbf{k}\cdot\mathbf{x}_{2}\right)}e^{\frac{i}{\hbar}\left(E_{k'}t_{1}-\mathbf{k}'\cdot\mathbf{x}_{1}\right)} \langle \Omega |b_{\mathbf{p}_{f}}b_{\mathbf{k}'}^{\dagger}\psi_{I}\left(x_{1}\right)\psi_{I}^{\dagger}\left(x_{2}\right)b_{\mathbf{k}}b_{\mathbf{p}_{i}}^{\dagger} |\Omega \rangle \nonumber \\
& = &
\frac{1}{L^{3}}\,e^{-\frac{i}{\hbar}\left(E_{i}t_{2}-\mathbf{p}_{i}\cdot\mathbf{x}_{2}\right)}e^{\frac{i}{\hbar}\left(E_{f}t_{1}-\mathbf{p}_{f}\cdot\mathbf{x}_{1}\right)}\langle \Omega |\psi_{I}\left(x_{1}\right)\psi_{I}^{\dagger}\left(x_{2}\right) |\Omega \rangle \nonumber \\
& = &
\frac{1}{L^{6}}\;e^{-\frac{i}{\hbar}\left(E_{i}t_{2}-\mathbf{p}_{i}\cdot\mathbf{x}_{2}\right)}e^{\frac{i}{\hbar}\left(E_{f}t_{1}-\mathbf{p}_{f}\cdot\mathbf{x}_{1}\right)}\sum_{\mathbf{k}}e^{-\frac{i}{\hbar}\left(E_{k}t_{1}-\mathbf{k}\cdot\mathbf{x}_{1}\right)}e^{\frac{i}{\hbar}\left(E_{k}t_{2}-\mathbf{k}\cdot\mathbf{x}_{2}\right)},
\end{eqnarray}
and thus we obtain:
\begin{eqnarray} \label{eq:sdktbd}
T^{\left(2\right)}\left(\mathbf{p}_{f};\mathbf{p}_{i};t\right) & = &
-\gamma_{m}\int_{0}^{t}dt_{1}\int_{0}^{t_{1}}dt_{2}\frac{1}{L^{6}}\int d\mathbf{x}_{1}\int d\mathbf{x}_{2}w\left(x_{1}\right)w\left(x_{2}\right)e^{-\frac{i}{\hbar}\left(E_{i}t_{2}-\mathbf{p}_{i}\cdot\mathbf{x}_{2}\right)}e^{\frac{i}{\hbar}\left(E_{f}t_{1}-\mathbf{p}_{f}\cdot\mathbf{x}_{1}\right)} \nonumber \\
& & \cdot \sum_{\mathbf{k}}e^{-\frac{i}{\hbar}\left(E_{k}t_{1}-\mathbf{k}\cdot\mathbf{x}_{1}\right)}e^{\frac{i}{\hbar}\left(E_{k}t_{2}-\mathbf{k}\cdot\mathbf{x}_{2}\right)}.
\end{eqnarray}

\subsection{Computation of the Transition Probability}

The next step is to compute the transition probability $P_{kj}\left(\mathbf{p}_{i};t\right)$, defined in Eq.~\eqref{eq:sdgsetj}; now we reintroduce the mass-index $k$ and $j$. In the interaction picture, it reads:
\begin{equation}
P_{kj}\left(\mathbf{p}_{i};t\right) = \sum_{\mathbf{p}_{f}}e^{\frac{i}{\hbar}\left(E_{f}^{\left(k\right)}-E_{f}^{\left(j\right)}\right)t}\mathbb{E}\left[T_{Ik}^{*}\left(\mathbf{p}_{f};\mathbf{p}_{i};t\right)T_{Ij}\left(\mathbf{p}_{f};\mathbf{p}_{i};t\right)\right]
\;, \end{equation}
where $T_{Ij}\left(\mathbf{p}_{f};\mathbf{p}_{i};t\right)$ has been defined in Eq.~\eqref{eq:dfgfdgfwe}.
% \begin{eqnarray}
% T_{Ij}\left(\mathbf{p}_{f};\mathbf{p}_{i};t\right) & \equiv & T_{j}^{\left(0\right)}\left(\mathbf{p}_{f};\mathbf{p}_{i};t\right)+T_{j}^{\left(1\right)}\left(\mathbf{p}_{f};\mathbf{p}_{i};t\right)+T_{j}^{\left(2\right)}\left(\mathbf{p}_{f};\mathbf{p}_{i};t\right) \nonumber \\
% & = &\delta_{\mathbf{p}_{f},\mathbf{p}_{i}}+\underset{\textrm{1 noise }}{\underbrace{T_{j}^{\left(1\right)}\left(\mathbf{p}_{f};\mathbf{p}_{i};t\right)}}+\underset{\textrm{2 noises}}{\underbrace{T_{j}^{\left(2\right)}\left(\mathbf{p}_{f};\mathbf{p}_{i};t\right)}}\;.
% \end{eqnarray}
When average over the noise, the non-zero terms are those which contain an even number of noises (in the graphical representation in terms of Feynman diagrams, only products of diagrams having an even number of dotted lines). Keeping terms only up to second order, we have:
\begin{eqnarray}
\mathbb{E}\left[T_{Ik}^{*}\left(\mathbf{p}_{f},\mathbf{p}_{i},t\right)T_{Ij}\left(\mathbf{p}_{f},\mathbf{p}_{i},t\right)\right] & = & \delta_{\mathbf{p}_{f},\mathbf{p}_{i}}+\delta_{\mathbf{p}_{f},\mathbf{p}_{i}}\mathbb{E}\left[T_{j}^{\left(2\right)}\left(\mathbf{p}_{f};\mathbf{p}_{i};t\right)\right]+\mathbb{E}\left[T_{k}^{\left(2\right)*}\left(\mathbf{p}_{f};\mathbf{p}_{i};t\right)\right]\delta_{\mathbf{p}_{f},\mathbf{p}_{i}} \nonumber \\
& + & \mathbb{E}\left[T_{k}^{\left(1\right)*}\left(\mathbf{p}_{f};\mathbf{p}_{i};t\right)T_{j}^{\left(1\right)}\left(\mathbf{p}_{f};\mathbf{p}_{i};t\right)\right]\;.
\end{eqnarray}
Therefore we get:
\begin{equation}\label{eq:P_kj}
P_{kj}\left(\mathbf{p}_{i};t\right)=e^{\frac{i}{\hbar}\left(E_{i}^{\left(k\right)}-E_{i}^{\left(j\right)}\right)t}\left[1+I_{j}^{\left(2\right)}\left(\mathbf{p}_{i};t\right)+I_{k}^{\left(2\right)*}\left(\mathbf{p}_{i};t\right)+I_{jk}^{\left(1\right)}\left(\mathbf{p}_{i};t\right)\right]
\end{equation}
where we have introduced the abbreviations:
\begin{eqnarray}
&&I_{j}^{\left(2\right)}\left(\mathbf{p}_{i};t\right)\equiv\mathbb{E}\left[T_{j}^{\left(2\right)}\left(\mathbf{p}_{i};\mathbf{p}_{i};t\right)\right]\;,\nonumber\\
&&I_{jk}^{\left(1\right)}\left(\mathbf{p}_{i};t\right)\equiv e^{\frac{i}{\hbar}\left(E_{i}^{\left(j\right)}-E_{i}^{\left(k\right)}\right)t}\sum_{\mathbf{p}_{f}}e^{\frac{i}{\hbar}\left(E_{f}^{\left(k\right)}-E_{f}^{\left(j\right)}\right)t}\,\mathbb{E}\left[T_{k}^{\left(1\right)*}\left(\mathbf{p}_{f};\mathbf{p}_{i};t\right)T_{j}^{\left(1\right)}\left(\mathbf{p}_{f};\mathbf{p}_{i};t\right)\right].
\end{eqnarray}
These two terms are explicitly computed in Appendix~B and~C. Here we report only the final result, which is:
\begin{equation}
I_{jk}^{\left(1\right)}\left(\mathbf{p}_{i};t\right)=\sqrt{\gamma_{m_{j}}\gamma_{m_{k}}}\left[2\int_{0}^{t}dsf\left(s\right)\left(t-s\right)\right]\frac{1}{\left(2\pi\right)^{3}}\frac{\pi^{3/2}}{r_{C}^{3}}
\end{equation}
and
\begin{equation}
I_{j}^{\left(2\right)}\left(\mathbf{p}_{i};t\right)=-\gamma_{m_{j}}\left[\int_{0}^{t}dsf\left(s\right)\left(t-s\right)\right]\frac{1}{\left(2\pi\right)^{3}}\frac{\pi^{3/2}}{r_{C}^{3}}\,,
\end{equation}
where $f(s)$ is the temporal correlation function of the noise.
Inserting the expression for $I_{jk}^{\left(1\right)}\left(\mathbf{p}_{i};t\right)$ and $I_{j}^{\left(2\right)}\left(\mathbf{p}_{i};t\right)$ written above in Eq.~\eqref{eq:P_kj} one obtains:

\begin{eqnarray}\label{eq:P_kjfinal}
P_{kj}\left(\mathbf{p}_{i};t\right) & = & e^{\frac{i}{\hbar}\left(E_{i}^{\left(k\right)}-E_{i}^{\left(j\right)}\right)t}\left\{ 1-\frac{\left(\sqrt{\gamma_{m_{j}}}-\sqrt{\gamma_{m_{k}}}\right)^{2}}{8\pi^{3/2}r_{C}^{3}}\left[\int_{0}^{t}dsf\left(s\right)\left(t-s\right)\right]\right\} \nonumber \\
& = & e^{\frac{i}{\hbar}\left(E_{i}^{\left(k\right)}-E_{i}^{\left(j\right)}\right)t}\left\{ 1-\frac{\gamma\left(m_{j}-m_{k}\right)^{2}}{8\pi^{3/2}r_{C}^{3}{m^{2}_{0}}}\left[\int_{0}^{t}dsf\left(s\right)\left(t-s\right)\right]\right\}\;,
\end{eqnarray}
where in the second line we use the definition Eq.~\eqref{eq:gamma}. Finally, the transition probability, Eq.~\eqref{eq:sdgsetj}, is computed to be:
\begin{eqnarray}
P_{K^{0}\rightarrow\bar{K}^{0}}\left(\mathbf{p}_{i}\right) & = & \frac{1}{4}\left[P_{SS}\left(\mathbf{p}_{i};t\right)-P_{LS}\left(\mathbf{p}_{i};t\right)-P_{SL}\left(\mathbf{p}_{i};t\right)+P_{LL}\left(\mathbf{p}_{i};t\right)\right]= \nonumber \\
& = & \frac{1}{2}\biggl[ 1-\cos\left[\frac{1}{\hbar}\left(E_{i}^{\left(L\right)}-E_{i}^{\left(S\right)}\right)t\right]\left\{ 1-\frac{\gamma\left(m_{j}-m_{k}\right)^{2}}{8\pi^{3/2}r_{C}^{3}{m^{2}_{0}}}\left[\int_{0}^{t}dsf\left(s\right)\left(t-s\right)\right]\right\}\biggr]\;.
\end{eqnarray}

In analogous way we can find the probability that a $K^{0}$ remains
$K^{0}$ with the time evolution:
\begin{eqnarray}
P_{K^{0}\rightarrow K^{0}}\left(\mathbf{p}_{i}\right) & = & \frac{1}{4}\left[P_{SS}\left(\mathbf{p}_{i};t\right)+P_{LS}\left(\mathbf{p}_{i};t\right)+P_{SL}\left(\mathbf{p}_{i};t\right)+P_{LL}\left(\mathbf{p}_{i};t\right)\right]= \nonumber \\
& = &  \frac{1}{2}\biggl[ 1+\cos\left[\frac{1}{\hbar}\left(E_{i}^{\left(L\right)}-E_{i}^{\left(S\right)}\right)t\right]\left\{ 1-\frac{\gamma\left(m_{j}-m_{k}\right)^{2}}{8\pi^{3/2}r_{C}^{3}{m^{2}_{0}}}\left[\int_{0}^{t}dsf\left(s\right)\left(t-s\right)\right]\right\}\biggr].
\end{eqnarray}
The computations starting from the anti-kaon are completely analogous, since we neglected $\mathcal{CP}$ violation.
Let us list here on two observations:

\begin{enumerate}
\item[(1)] The probabilities $P_{K^{0}\rightarrow K^{0}}\left(\mathbf{p}_{i}\right)$
and $P_{K^{0}\rightarrow\bar{K}^{0}}\left(\mathbf{p}_{i}\right)$
sum to $1$, which means that there are no particle losses.

\item[(2)] The factor after the cosine is due to the noise present in this
model, i.e. if there is no noise ($\gamma=0)$ the square bracket gives
$1$ and arrive at the standard oscillations formula\footnote{In some of these works, inside the cosine, only the mass difference appears. This is the same result as ours, if one makes the approximation (good at the non-relativistic level) $E_{i}^{\left(j\right)}=m_{j}c^{2}+\frac{\mathbf{p}_{i}^{2}}{2m_{j}}\simeq m_{j}c^{2}$.}~\cite{BDH1,BenFlorRom}.
\end{enumerate}
Since the effect of the noise is usually very similar to the
one given by the decoherence~\cite{BDH1,BenFlorRom}, and it is well know from the literature
that in such cases decoherence dumps the oscillation with an exponential function,
we can think at the term inside the square bracket as the first term
of the series of an exponential function.
Therefore, we can guess that the exact result
for the probability transition from a kaon to an anti-kaon could reasonably
be:
\begin{equation}\label{finalsingleprobability}
P_{K^{0}\rightarrow\bar{K}^{0}}\left(\mathbf{p}_{i}\right)=\frac{1}{2}\biggl\lbrace 1-\cos\left[\frac{1}{\hbar}\left(E_{i}^{\left(L\right)}-E_{i}^{\left(S\right)}\right)t\right]\exp
\left\{-\frac{\gamma\left(m_{j}-m_{k}\right)^{2}}{8\pi^{3/2}r_{C}^{3}{m^{2}_{0}}}\left[\int_{0}^{t}dsf\left(s\right)\left(t-s\right)\right]\right\}\biggr\rbrace\;.
\end{equation}

To conclude the computation on a single particle, we can include the decay of the particle adding to the free Hamiltonian
$H_{S}$ an imaginary term:
\begin{equation}
H_{jS}\longrightarrow H_{jS}-\frac{i}{2}\Gamma_{j}.
\end{equation}
This changes the previous computation by sending $E_{i}^{\left(j\right)}\longrightarrow E_{i}^{\left(j\right)}-\frac{i}{2}\Gamma_{j}$ and one can easily notice that the only change consists in multiplying each $P_{kj}\left(\mathbf{p}_{i};t\right)$ with the exponential function $e^{-\frac{\Gamma_{k}+\Gamma_{j}}{2\hbar}t}$ in order to take the decay into account. Thus we obtain the final result:
\begin{equation}\label{finalsingleprobability2}
P_{K^{0}\rightarrow\bar{K}^{0}}\left(\mathbf{p}_{i}\right)=\frac{1}{4}\biggl\lbrace e^{-\frac{\Gamma_{L}}{\hbar}t}+e^{-\frac{\Gamma_{S}}{\hbar}t}-
2\cos\left[\frac{1}{\hbar}\left(E_{i}^{\left(L\right)}-E_{i}^{\left(S\right)}\right)t\right]\cdot e^{-\frac{\Gamma_{L}+\Gamma_{S}}{2\hbar}t}\cdot
\underbrace{e^{-\frac{\gamma\left(m_{S}-m_{L}\right)^{2}}{16\pi^{3/2}r_{C}^{3}m_{0}^{2}}t}}_{\textrm{effect due to CLS model}}\biggr\rbrace
\end{equation}
where now we have assumed that the noise is white in time, i.e. $f(s) = \delta(s)$. This is the standard situation with the CSL model.
Similar results hold for the other transition probabilities.

\section{The Collapse Model for Two Particle States}

At the DA$\Phi$NE collider~\cite{Kloe1,DiDomenico,DiDomenico2,HiesmayrKLOE} neutral kaons are copiously produced in an entangled antisymmetric state:
\begin{equation}\label{antiysmmetricstate}
\left|I\right\rangle =\frac{\left|K_{S}K_{L}\right\rangle -\left|K_{L}K_{S}\right\rangle }{\sqrt{2}}=\frac{\left|\bar{K}^{0}K^{0}\right\rangle -\left|K^{0}\bar{K}^{0}\right\rangle }{\sqrt{2}}\;.
\end{equation}
We want here to investigate how the mass-proportional CSL model changes the time evolution of entangled states. In order to obtain a more general
result, we perform the computation for an arbitrary two-particle
state. One particle evolves to the left hand side and the other kaon evolves to the right hand side, for which they need the time $t_l, t_r$, respectively. It is convenient to use the mass basis $\left|K_{S}K_{S}\right\rangle ,\left|K_{S}K_{L}\right\rangle ,\left|K_{L}K_{S}\right\rangle ,\left|K_{L}K_{L}\right\rangle $,
where the state on the left hand side is a plane wave with momentum $-\mathbf{p}_{i}$ and the one on the right hand side is a
plane wave with momentum $\mathbf{p}_{i}$. In the Fock space framework, this means for example $\left|K_{S}K_{L}\right\rangle =a_{S}^{\dagger}\left(-\mathbf{p}_{i}\right)a_{K}^{\dagger}\left(\mathbf{p}_{i}\right)\left|\Omega\right\rangle $.
Since we are considering only states with two particles with
definite momentum $\pm\mathbf{p}_{i}$, the four states above
form a complete basis. So the generic initial state can be decomposed as:
\begin{equation}
\left|I\right\rangle =\sum_{j,k=S,L}\alpha_{jk}\left|K_{j}K_{k}\right\rangle \;\;\;\textrm{with}\;\;\;\sum_{j,k=S,L}\left|\alpha_{jk}\right|^{2}=1\;.
\end{equation}
Let us assume that we want to know the probability to find the left
particle at time $t_{l}$ in the state $\left|F_{l},\mathbf{p}_{l}\right\rangle =\sum_{m=S,L}\beta_{m}\left|K_{m},\mathbf{p}_{l}\right\rangle $
and the right particle at time $t_{r}$ in the state $\left|F_{r},\mathbf{p}_{r}\right\rangle =\sum_{n=S,L}\gamma_{n}\left|K_{n},\mathbf{p}_{r}\right\rangle $.
Here $\left|F_{l}\right\rangle $ and $\left|F_{r}\right\rangle $
can be mass or flavor eigenstates. In the end we have to sum over all $\mathbf{p}_{l}$
and all $\mathbf{p}_{r}$ since we are interested in a result independent from the
particular final momentum of the particles. We start by computing the amplitude:
\begin{eqnarray}
A\left(F_{l},\mathbf{p}_{l};F_{r},\mathbf{p}_{r}\right) & = & \left\langle F_{l},\mathbf{p}_{l};F_{r},\mathbf{p}_{r}\left|U_{\text{\tiny LEFT}}\left(t_{l}\right)\otimes U_{\text{\tiny RIGHT}}\left(t_{r}\right)\right|I\right\rangle= \nonumber \\
& = & \sum_{j,k=S,L}\alpha_{jk}\left\langle F_{l},\mathbf{p}_{l}\left|U_{j}\left(t_{l}\right)\right|K_{j},-\mathbf{p}_{i}\right\rangle \left\langle F_{r},\mathbf{p}_{r}\left|U_{k}\left(t_{r}\right)\right|K_{k},\mathbf{p}_{i}\right\rangle =\nonumber \\
& = & \sum_{j,k,m,n=S,L}\alpha_{jk}\beta_{m}^{*}\gamma_{n}^{*}\left\langle K_{m},\mathbf{p}_{l}\left|U_{j}\left(t_{l}\right)\right|K_{j},-\mathbf{p}_{i}\right\rangle \left\langle K_{n},\mathbf{p}_{r}\left|U_{k}\left(t_{r}\right)\right|K_{k},\mathbf{p}_{i}\right\rangle =\nonumber \\
& = & \sum_{j,k=S,L}\alpha_{jk}\beta_{j}^{*}\gamma_{k}^{*}\left\langle K_{j},\mathbf{p}_{l}\left|U_{j}\left(t_{l}\right)\right|K_{j},-\mathbf{p}_{i}\right\rangle \left\langle K_{k},\mathbf{p}_{r}\left|U_{k}\left(t_{r}\right)\right|K_{k},\mathbf{p}_{i}\right\rangle,
\end{eqnarray}
since it gives the probability of interest:
\begin{eqnarray}\label{eq:Pdouble}
P\left(F_{l};F_{r}\right) & \equiv & \sum_{\mathbf{p}_{l},\mathbf{p}_{r}}\mathbb{E}\left|A\left(F_{l},\mathbf{p}_{l};F_{r},\mathbf{p}_{r}\right)\right|^{2}=\sum_{j,k,j',k'=S,L}\alpha_{jk}\beta_{j}^{*}\gamma_{k}^{*}\alpha_{j'k'}^{*}\beta_{j'}\gamma_{k'} \nonumber \\
& \times & \mathbb{E} \left\{\left[\sum_{\mathbf{p}_{l}}\left\langle K_{j'},\mathbf{p}_{l}\left|U_{j'}\left(t_{l}\right)\right|K_{j'},-\mathbf{p}_{i}\right\rangle ^{*}\left\langle K_{j},\mathbf{p}_{l}\left|U_{j}\left(t_{l}\right)\right|K_{j},-\mathbf{p}_{i}\right\rangle \right]\right.\nonumber \\
& \times & \left.\left[\sum_{\mathbf{p}_{r}}\left\langle K_{k'},\mathbf{p}_{r}\left|U_{k'}\left(t_{r}\right)\right|K_{k'},\mathbf{p}_{i}\right\rangle ^{*}\left\langle K_{k},\mathbf{p}_{r}\left|U_{k}\left(t_{r}\right)\right|K_{k},\mathbf{p}_{i}\right\rangle \right]\right\}\;.
\end{eqnarray}

The noise average involves terms that are contained in the
left square bracket (the particle on the left hand side), or in the right square bracket (the particle on the right hand side) and
mixed terms where the correlation is taken between a piece from the
first square bracket and the other piece from the second one. Since
we work with plane waves all these different terms are of importance.
In the more realistic situation, however, where we assume to have initial
confined wave packets that propagate in opposite directions, it can
be show (see appendix~D) that the mixed terms give a negligible contribution compared
to the other ones. This considerably simplifies the computation to
\begin{eqnarray}
P\left(F_{l};F_{r}\right) & = & \sum_{\mathbf{p}_{l}}\mathbb{E}\left[\left\langle K_{j'},\mathbf{p}_{l}\left|U_{j'}\left(t_{l}\right)\right|K_{j'},-\mathbf{p}_{i}\right\rangle ^{*}\left\langle K_{j},\mathbf{p}_{l}\left|U_{j}\left(t_{l}\right)\right|K_{j},-\mathbf{p}_{i}\right\rangle \right] \nonumber \\
& \times & \sum_{\mathbf{p}_{r}}\left[\mathbb{E}\left\langle K_{k'},\mathbf{p}_{r}\left|U_{k'}\left(t_{r}\right)\right|K_{k'},\mathbf{p}_{i}\right\rangle ^{*}\left\langle K_{k},\mathbf{p}_{r}\left|U_{k}\left(t_{r}\right)\right|K_{k},\mathbf{p}_{i}\right\rangle \right].
\end{eqnarray}

The terms inside the square bracket are the same as those we computed in the previous sections for the single particle case, therefore we obtain a factorization of the probabilities, i.e.
\begin{equation}
P\left(F_{l};F_{r}\right)=\sum_{j,k,j',k'=S,L}\alpha_{jk}\beta_{j}^{*}\gamma_{k}^{*}\alpha_{j'k'}^{*}\beta_{j'}\gamma_{k'}\;
P_{j'j}\left(-\mathbf{p}_{i};t_{l}\right)\cdot P_{k'k}\left(\mathbf{p}_{i};t_{r}\right),
\end{equation}
with $P_{kj}\left(\mathbf{p}_{i};t\right)$ given by Eq.~\eqref{eq:P_kjfinal}.

For the antisymmetric initial state~\eqref{antiysmmetricstate} ($\alpha_{SL}=-\alpha_{LS}=\frac{1}{\sqrt{2}}$ else $\alpha_{ij}=0$) and choosing the final states, $\left|F_{l}\right\rangle =\left|K^{0}\right\rangle $ ($\beta_S=\beta_L=\frac{1}{\sqrt{2}}$)
and $\left|F_{r}\right\rangle =\left|K^{0}\right\rangle $ ($\gamma_S=\gamma_L=\frac{1}{\sqrt{2}}$), respectively, we have to compute
\begin{eqnarray}
P\left(K^{0};K^{0}\right)& = & \frac{1}{8}\left[P_{SS}\left(-\mathbf{p}_{i};t_{l}\right)\cdot P_{LL}\left(\mathbf{p}_{i};t_{r}\right)+P_{LS}\left(-\mathbf{p}_{i};t_{l}\right)\cdot P_{SL}\left(\mathbf{p}_{i};t_{r}\right)\right.\ \nonumber \\
& + & \left.\ P_{SL}\left(-\mathbf{p}_{i};t_{l}\right)\cdot P_{LS}\left(\mathbf{p}_{i};t_{r}\right)+P_{LL}\left(-\mathbf{p}_{i};t_{l}\right)\cdot P_{SS}\left(\mathbf{p}_{i};t_{r}\right)\right]\;.
\end{eqnarray}
where in the case of white noise field we have:
\begin{equation}
P_{kj}\left(\mathbf{p}_{i};t\right)=e^{-\frac{\Gamma_{k}+\Gamma_{j}}{2\hbar}t}e^{\frac{i}{\hbar}\left(E_{i}^{\left(j\right)}-E_{i}^{\left(k\right)}\right)t}\cdot
e^{-\frac{\gamma\left(m_{j}-m_{k}\right)^{2}}{16\pi^{3/2}r_{C}^{3}m_{0}^{2}}t}
\end{equation}
and thus:
\begin{eqnarray}\label{finaljointprobability}
P\left(K^{0};K^{0}\right) & = & \frac{1}{8}\left\{ e^{-\frac{\Gamma_S}{\hbar}t_{l}-\frac{\Gamma_L}{\hbar}t_{r}}+e^{-\frac{\Gamma_L}{\hbar}t_{l}-\frac{\Gamma_S}{\hbar}t_{r}}\right. \nonumber \\
& + & \left.2\cdot \cos\left[\frac{1}{\hbar}\left(E_{i}^{\left(S\right)}-E_{i}^{\left(L\right)}\right)\left(t_{r}-t_{l}\right)\right]
\cdot e^{-\frac{\Gamma_{L}+\Gamma_{S}}{2\hbar}\left(t_{l}+t_{r}\right)}\cdot \underbrace{e^{-\frac{\gamma\left(m_{S}-m_{L}\right)^{2}}{16\pi^{3/2}r_{C}^{3}m_{0}^{2}}\left(t_{l}+t_{r}\right)}}_{\textrm{effect due to CLS model}}\right\}\;.
\end{eqnarray}
In the next section, we comment on the experimental implications of the above formula.

%%%%%%%%%%%%%%%%%%%%%%%%%%%%%%%%%%%%%%%%%%%%%%%%%%%%%%%%%%%%%%%%%%%%%%%%%%%%%%%%%%%%%%%%%%%%%%%%%%%%%%%%%%%%%%%%%%%%%%%
\section{Estimation Of the Effect Of the Collapse On Single and Entangled Kaons \& Connections to Other Models}

%To estimate the damping factor we choose to study the white noise case $f\left(s\right)=\delta\left(s\right)$ and so in the previous
%formula all the square bracket in the exponential factor becomes just $t/2$. In such case the transition probability is:
%\begin{equation}
%P_{K^{0}\rightarrow\bar{K}^{0}}\left(\mathbf{p}_{i}\right)=\frac{1}{2}-\frac{1}{2}\cos\left[\frac{1}{\hbar}\left(E_{i}^{\left(L\right)}-E_{i}^{\left(S\right)}\right)t\right]\exp\left[-\frac{\gamma\left(m_{S}-m_{L}\right)^{2}}{16\pi^{3/2}r_{c}^{3}m_{0}^{2}}t\right].
%\end{equation}

The effect of the collapse is the introduction of an additional damping of the interference term. This effect is also observed in the case one assumes that during the time evolution of the single or two-particle state the kaon or kaons interact with an environment in the mass basis $\{K_S,K_L\}$, which was investigated by the authors of Ref.~\cite{BDH1}. They treated the neutral kaons as an open quantum system with Markovian interactions (for an overview on open quantum system see e.g. \cite{BPopenquantum}) and their dynamics is given by the Liouville-von Neumann equation with an additional Lindblad term:
\begin{eqnarray}{\cal D}[\rho]=\frac{1}{2}\sum  (L_j^\dagger
L_j \rho+\rho\;L_j^\dagger L_j-2  L_j \rho
L_j^\dagger)\end{eqnarray}
which gives rise to the well known Lindbald equation~\cite{Lindblad,GoriniKossakowskiSudarshan}:
\begin{eqnarray}\label{masterequation}
\frac{d}{dt}\rho&=&-i\, H\; \rho+i\,\rho\; H^\dagger-\cal{D}[\mathbf{\rho}]\;.
\end{eqnarray}
They further assumed that the interaction occurs in the mass basis. For the single particle case one chooses the Lindbald generators to be $L_{S,L}=\sqrt{\Lambda_{\textrm{single}}} |K_{S,L}\rangle\langle K_{S,L}|$ where $\Lambda_{\textrm{single}}$ is the strength of the interaction. The solution of the components of the density matrix $\rho=\sum_{i,j=S,L}\rho_{ij}(t) |K_{i}\rangle\langle K_{j}|$ is given by:
\begin{eqnarray}
\rho_{ij}(t)=\rho_{ij}(0)\cdot e^{-\Gamma_i t \delta_{i,j}+(1-\delta_{i,j}) (i \frac{c^2}{\hbar}(m_i-m_j) t-\frac{\Gamma_S+\Gamma_L}{2 \hbar} t-\Lambda_{\textrm{single}}\cdot t)}\;,
\end{eqnarray}
thus, as we expect, only the off-diagonal terms are affected by the interaction with the environment. In particular, the probability to find an antikaon when a kaon state was produced at time $t=0$ becomes:
\begin{eqnarray}\label{probdeco}
P_{K^{0}\rightarrow\bar{K}^{0}}(t)&=&\frac{1}{4}\biggl\lbrace e^{-\frac{\Gamma_S}{\hbar} t}+e^{-\frac{\Gamma_L}{\hbar} t}- 2 \cos\left[\frac{c^2}{\hbar}\left(m_S-m_L\right) t\right]\; e^{-\frac{\Gamma_S+\Gamma_L}{2 \hbar} t}\cdot \underbrace{e^{-\Lambda_{\textrm{single}}\cdot t}}_{\textrm{decoherence effect}}\biggr\rbrace
\end{eqnarray}
which is of the same structure as the result, Eq.~(\ref{finalsingleprobability2}), based on the mass dependent CLS model. In the two particle case we have a similar behaviour if the two Lindbald generators are chosen to be $L_{1}=\sqrt{\Lambda_{\textrm{two-particle}}} |e_1\rangle\langle e_1| $ and $L_{2}=\sqrt{\Lambda_{\textrm{two-particle}}}|e_2\rangle\langle e_2|$ with $|e_1\rangle=|K_{S}\rangle\otimes|K_{L}\rangle$ and $|e_2\rangle=|K_{L}\rangle\otimes|K_{S}\rangle$. Then for the initial antisymmetric state  $\rho=|\psi^-\rangle\langle\psi^-|$ with $|\psi^-\rangle=\frac{1}{\sqrt{2}}\{|e_1\rangle-|e_2\rangle\}$ we obtain the following time dependent density matrix before one kaon decayed:
\begin{eqnarray}
\rho(t)&=& \frac{1}{2} e^{-\frac{\Gamma_S+\Gamma_L}{2\hbar} t}\biggl\lbrace |e_1\rangle\langle e_1|+|e_2\rangle\langle e_2|-e^{-\Lambda_{\textrm{two-particle}}\cdot t}(|e_1\rangle\langle e_2|+|e_2\rangle\langle e_1|)\biggr\rbrace\;.
\end{eqnarray}
Assuming that after one particle decayed no further decoherence effect occurs, namely only the two-particle state interacts with the environment, then we obtain the following joint probability~\cite{BDH1}:
\begin{eqnarray}
P\left(K^{0}, t_l;K^{0} t_r\right) & = & \frac{1}{8}\left\{ e^{-\frac{\Gamma_S}{\hbar}t_{l}-\frac{\Gamma_L}{\hbar}t_{r}}+e^{-\frac{\Gamma_L}{\hbar}t_{l}-\frac{\Gamma_S}{\hbar}t_{r}}\right. \nonumber \\
& + & \left.2\cdot \cos\left[\frac{c^2}{\hbar}\left(m_S-m_L\right)\left(t_{r}-t_{l}\right)\right]
\cdot e^{-\frac{\Gamma_{L}+\Gamma_{S}}{2\hbar}\left(t_{l}+t_{r}\right)}\cdot \underbrace{e^{-\Lambda_{\textrm{two-particle}}\cdot\min\left\{t_{l},t_{r}\right\}}}_{\textrm{decoherence effect}}\right\}
\end{eqnarray}
that has the same structure as the result obtained from the CSL model, Eq.~(\ref{finaljointprobability}), except that here the damping is not sensitive to the sum of the times, but on the time value corresponding to the first measured kaon. The underlying philosophy is the assumption that an entangled two-particle state has a single time evolution and interacts as a whole system with the environment. When one kaon is measured or decays there is no further interaction with the environment or the single particle state interacts via another interaction with the environment. Thus the longer an entangled pair of kaons survives, the more it interacts with the environment, namely the effect increases. So far, though investigations for kaons and B-mesons~\cite{Richter} have been tried, no effect has been found
%it was not possible to retrieve any time resolution
from the currently available data.

Another option is to assume that both kaons, though being entangled, separately interact with the environment. This is the case for the considered CSL model, because it assumes that the contribution from the collapse is the sum of contributions to the Hamiltonian of the short-lived state and the long-lived state.

Let us now compare with data of experiments and estimate the effects. There exist precision experiments for neutral K-mesons and neutral B-mesons in the antisymmetric entangled state $|\psi^-\rangle$ that investigate bounds on these possible decoherence effects or differently stated on the stability of nonlocal correlations. They consider the joint probabilities and model them by one single phenomenological parameter $\zeta$ multiplying the interference effect, i.e.:
\begin{eqnarray}\label{decoherenceprobability}
P\left(K^{0};K^{0}\right) & = & \frac{1}{8}\left\{ e^{-\frac{\Gamma_S}{\hbar}t_{l}-\frac{\Gamma_L}{\hbar}t_{r}}+e^{-\frac{\Gamma_L}{\hbar}t_{l}-\frac{\Gamma_S}{\hbar}t_{r}}\right. \nonumber \\
& + & \left.2\cdot \cos\left[\frac{1}{\hbar}\left(E_{i}^{\left(S\right)}-E_{i}^{\left(L\right)}\right)\left(t_{r}-t_{l}\right)\right]
\cdot e^{-\frac{\Gamma_{L}+\Gamma_{S}}{2\hbar}\left(t_{l}+t_{r}\right)}\cdot (1-\zeta)\right\}\;.
\end{eqnarray}
For data they get the following numbers. The authors of Ref.~\cite{BGH1} analyzed the 1999-CPLEAR experiment~\cite{CPLEAR} and found the following upper bounds if the decoherence effect is assumed to be in the mass basis (as discussed above) averaged over both measured time configurations:
\begin{eqnarray}
\bar{\zeta}=0.13\big\lbrace^{+0.16}_{-0.15}
\end{eqnarray}
from which we estimate via $1-\bar\zeta\equiv e^{-\Lambda_{\textrm{two-particle}}\cdot\min\left\{t_{l},t_{r}\right\}}$ the upper bound of effects coming from possible interactions with the environment to be (C.L.\dots confidence level):
\begin{eqnarray}
%\bar\Lambda_{\textrm{two-particle}}&=&0.28\cdot 10^9\; s^{-1}\qquad\textrm{or}\qquad \tilde\Lambda_{\textrm{two-particle}}\;:=\;\frac{\bar\Lambda_{\textrm{two-particle}}}{\Gamma_S}\;=\;0.25\;.
&&\Lambda_{\textrm{two-particle}}\;\leq\;8.8\cdot 10^9\; s^{-1}\quad\textrm{at}\quad 90\%\;\textrm{C.L.}\qquad\textrm{or}\nonumber\\
&&\tilde\Lambda_{\textrm{two-particle}}\;:=\;\frac{\Lambda_{\textrm{two-particle}}}{\Gamma_S}\;\leq\;0.79\quad\textrm{at}\quad 90\%\; \textrm{C.L.}\;.
\end{eqnarray}
Note that we have only in this case a one-to-one correspondence of the $\zeta$ approach and the $\Lambda_{\textrm{two-particle}}$ approach since in both measured time configurations the minimal time was the same.

Let us compare this result with  the theoretical prediction given by the mass-proportional CSL model. Using the stronger value suggested by Adler for $\gamma$, namely  $\gamma=10^{-22}\textrm{cm}^{3}\textrm{s}^{-1}$~\cite{adlerphoto}, and recalling that $r_{C}=10^{-5}\textrm{cm}$,
$\left|m_{S}-m_{L}\right|\simeq 3.5\cdot 10^{-12}\textrm{MeV}/c^{2}$~\cite{ParticleDataBook},
 and $m_{0}\simeq9.4\cdot 10^{2}\textrm{MeV}/c^{2}$~\cite{ParticleDataBook}, we obtain:
\begin{eqnarray}
\bar\Lambda_{\textrm{CSL}}&:=&\frac{\gamma\left(m_{S}-m_{L}\right)^{2}}{16\pi^{3/2}r_{C}^{3}m_{0}^{2}}%=\frac{3,5^{2}}{16\times\pi^{3/2}\times9,4^{2}}10^{-22-24+15-4}
=1.5\times10^{-38}\textrm{s}^{-1}\qquad\textrm{or}\qquad \tilde\Lambda_{\textrm{CSL}}\;:=\;\frac{\bar\Lambda_{\textrm{CSL}}}{\Gamma_S}\;=\;1.3\cdot 10^{-48}\;.
\end{eqnarray}
Thus the expected effect is by many orders smaller than the sensitivity given by the CPLEAR experiment~\cite{CPLEAR}.

The KLOE collaboration~\cite{DiDomenico2,DiDomenico,Kloe1} exploring the antisymmetric entangled kaon state produced at the $\Phi$ resonance has also investigated possible decoherence effect. It is a very clean experiment to look for small effects. The above idea of the $\zeta$-parameter~\cite{BGH1} was investigated---differently from the above described experiment---for the final states $\pi^+\pi^-$ on both sides. Moreover, the decoherence strength $\zeta$ was achieved by integration over the natural distribution of the time differences. It gives so far the highest precision due to a $\mathcal{CP}$ suppressing mechanism, i.e. the value is:
\begin{eqnarray}
\bar{\zeta}_{\pi^+,\pi^-}=0.003\pm0.018_\textrm{stat}\pm 0.006_\textrm{syst}\;.
\end{eqnarray}
Collapse models apply to all systems, and for small times one has: $\zeta\approx \Lambda_{\textrm{CSL}}\cdot t$. From that we can estimate that the sensitivity of the above experiment, namely the statistical sensitivity, the detector resolutions and the performance of the chosen setup, is not high enough to show the possible effects of the CSL model. Certainly, one has in this case to do the calculation including $\mathcal{CP}$ violation, but the final conclusion should not change significantly.

In summary, the experimental situation at the DA$\Phi$NE collider, where the experiments of the KLOE collaboration are performed, provides a very clean environment in the sense that environmental effects can be kept low, however, as the above averaged value of $\zeta$ shows the limiting errors are the statistical one and the systematic one given by the present technology. Looking for better suited time measurements (no average), better suited observables and more statistics might considerably
%can considerable
enhance the sensitivity.

Let us now discuss the other meson systems. For all mesons types except the K-mesons the decay widths are in good approximation equal (see appendix~A), $\Gamma_L\simeq\Gamma_H\simeq\Gamma$, thus  the probabilities to find on both sides the same flavor eigenstates $M^0=B,D,B_s$ at different times is given by (compare with Eq.~(\ref{decoherenceprobability})):
\begin{eqnarray}
P\left(M^{0}, t_l;M^{0}, t_r \right) & = & \frac{e^{-\frac{\Gamma}{\hbar}(t_{l}+t_{r})}}{2}\left\{1+ \cos\left[\frac{1}{\hbar}\left(E_{i}^{\left(S\right)}-E_{i}^{\left(L\right)}\right)\left(t_{r}-t_{l}\right)\right]\right\}\;,
\end{eqnarray}
hence the effect of the decay property is only an overall effect.
Let us compute the effect predicted by the CSL model using the data of \cite{ParticleDataBook} for these mesons:
\begin{eqnarray}
\bar\Lambda_{\textrm{CSL},B\textrm{-meson}}&:=&\frac{\gamma\left(m_{H}-m_{L}\right)^{2}}{16\pi^{3/2}r_{C}^{3}m_{0}^{2}}
=1.4\times10^{-34}\textrm{s}^{-1}\qquad\textrm{or}\qquad \tilde\Lambda_{\textrm{CSL}}\;:=\;\frac{\bar\Lambda_{\textrm{CSL},B\textrm{-meson}}}{\Gamma}\;=\;2.1\cdot 10^{-46}\;,\nonumber\\
\bar\Lambda_{\textrm{CSL},B_s\textrm{-meson}}&:=&\frac{\gamma\left(m_{H}-m_{L}\right)^{2}}{16\pi^{3/2}r_{C}^{3}m_{0}^{2}}
=1.7\times10^{-31}\textrm{s}^{-1}\qquad\textrm{or}\qquad \tilde\Lambda_{\textrm{CSL}}\;:=\;\frac{\bar\Lambda_{\textrm{CSL},B_s\textrm{-meson}}}{\Gamma}\;=\;2.6\cdot 10^{-42}\;,\nonumber\\
\bar\Lambda_{\textrm{CSL,D-meson}}&:=&\frac{\gamma\left(m_{H}-m_{L}\right)^{2}}{16\pi^{3/2}r_{C}^{3}m_{0}^{2}}
=3.2\times10^{-37}\textrm{s}^{-1}\qquad\textrm{or}\qquad \tilde\Lambda_{\textrm{CSL}}\;:=\;\frac{\bar\Lambda_{\textrm{CSL},\textrm{D-meson}}}{\Gamma}\;=\;1.3\cdot 10^{-49}\;.
\end{eqnarray}
We observe that the effect of the collapse compared with the mean time is the best for the $B_s$-meson. For B-mesons estimates for $\zeta^{\textrm{B-mesons}}$ exists~\cite{Appollo,BertlmannGrimus,Richter,Yabsley}, they are again far from the needed accuracy. Last but no least let us mention that authors discuss possible decoherence effects arising from different effects, e.g. the author of Ref.~\cite{Mavromatos2} discusses decoherence arising from the presence of dark energy in the universe and concludes that a possible decoherence effects due to quantum gravity strongly depends on the details of the structure of the quantum foam.

\section{Conclusions}

%{\bf CATALINA, ANTONIO, ANGELO, SANDRO, PLEASE EXTEND THE CONCLUSIONS?}

%Various experiments suggest that neutral mesons, particles that consist of a quark and an antiquark, are states that are superpositions of two different mass eigenstates.
In this paper we addressed the question whether models of spontaneous wave function collapse, which modify the standard quantum evolution by adding non-linear and stochastic terms to the Schr\"odinger equation,  are testable, or not, for single and even entangled
neutral meson systems in High Energy Physics.
If spontaneous collapses occur in Nature, we expect they would affect these systems and their flavor
oscillation mechanism in a non trivial way.
%
%Certainly, if spontaneous collapses occur in Nature, they would affect these systems at different energy scales, however, since they are superpositions of mass eigenstates they are uniquely affected by the known collapse models.

We have focussed our analysis on the mass-proportional CSL (Continuous Spontaneous Localization) model~\cite{Cslmass,Fu} and assumed that the contributions of the two different mass eigenstates are independently added to the Hamiltonian. With that we  connected the spatiotemporal CSL collapse mechanism with the two dimensional dynamics describing the flavor oscillations. In order to perform the lengthly and involved computation we had to use different methods. First, we replaced the collapse noise with an imaginary noise that has been shown to produce all the experimentally testable predictions of the CSL model. Next, the noise has been treated as a small perturbation.  Moving to the interaction picture we derived order by order via Dyson series the transition amplitudes. Via them we obtained the probabilities of interest for single and bipartite systems. In the very last step we assumed that the noise is white in time. Consequently, via our computation one can in future also consider the effects of different noise correlation functions.

The result of the computed probabilities within the CSL model shows that the oscillation terms are affected by an exponential damping. It is sensitive to the mass difference squared and the two phenomenological parameters of the CSL model, which, if the theory is taken seriously, acquire the status of new constants of nature. They are currently fixed to certain numbers~\cite{adlerphoto} %to be in agreement
which are compatible with the results of
all current experiments.
We computed also the case of bipartite entangled systems where we found that the effect of CSL factorizes, i.e. each meson seems to be affected separately and independently of the initial state they share.

We compared the CSL model with other models~\cite{BGH1,BDH1} that investigate possible decoherence effects or the stability of nonlocal correlations and that have been investigated at accelerator facilities~\cite{CPLEAR,DiDomenico2,DiDomenico,Kloe1}.
We want to note here, that the upper bound on the $\zeta$ parameter published by the KLOE collaboration gives us an estimate on the magnitude of the current sensitivity of the experiments.
Our overall conclusion is that the possible collapse effect on these mass-superposed states is too small to be seen by the up to date performed experiments.
A dedicated experiment, choosing other more suitable observables, times and with more statistics might enhance the
%bounds on the CSL parameters
sensitivity.

%However we do not expect it could in any way reach the sensitivity to detect the effects predicted by the CSL collapse model.
%
%Our overall conclusion is that the possible collapse effect on these mass-superposed states is too small to be seen with the current experiments, however, there is no dedicated experiment so far. The main point is that all existing experiments do not give a time resolution, thus there is no one-to-one correspondence between the theoretical predicted probabilities and experiments. We want also to note here, that the upper bound on the $\zeta$ parameter published by the KLOE collaboration gives us an estimate on the magnitude of the current sensitivity of the experiment but its statistical meaning or confidence level is difficult to estimate. Therefore, a dedicated experiment, choosing proper observables, times and more statistics could enhance the bound considerably.

Last but not least, we want to compare the results of mesons with the ones of neutrinos (for a detailed discussion see Refs.~\cite{neutrino1,neutrino}). In that case one has to connect the spatiotemporal propagation with the three dimensional Hilbert space of the lepton flavor oscillations. Since the mass is much lower than in the meson case one expects that the collapse effect is also considerably smaller and thus harder to detect, however, differently to kaons the control of the environmental effects is not possible since it is given mainly by interaction with other neutrinos while traveling through our universe.

\section*{Acknowledgements}

All authors would like to thank the COST Action MP1006  ``Fundamental Problems in Quantum Physics". The project is partly funded from the SoMoPro programme. Research of B.C.H. leading to these results has received a financial contribution from the European Community within the Seventh Framework Programme (FP/2007-2013) under Grant Agreement No. 229603. The research is also co-financed by the South Moravian Region.
A.B., C.C. and S.D. wish to thank S.L. Adler for many useful and enjoyable conversations on this topic. They also acknowledges partial financial support from MIUR (PRIN 2008), INFN and the John Templeton Foundation project `Quantum Physics and the Nature of Reality'. %B.C.H. wants also to thank the Austrian Science Fund (FWF) project P21947N16.

\section*{Appendix A: Kaon Phenomenology}

The phenomenology of oscillation and decay of meson-antimeson systems can
be described by nonrelativistic quantum mechanics effectively, because the
dynamics is depending on the observable hadrons rather than on the more
fundamental quarks. A quantum field theoretical calculation showing negligible
corrections can e.g. be found in Refs.~\cite{Capolupo1,Beuthe}.

A neutral meson $M_0$ is a bound state of quark and antiquark. As numerous experiments
have revealed the particle state $M_0$ and the antiparticle state $\bar M_0$ can decay into the same final
states, thus the system has to be handled as a two state system similar to
spin $\frac{1}{2}$ systems. In addition to being a decaying system these massive particles
show the phenomenon of flavor oscillation, i.e. an oscillation between matter
and antimatter occurs. If e.g. a neutral meson is produced at time $t = 0$ the
probability to find an antimeson at a later time is nonzero.

The most general time evolution for the two state system $M^0-\bar M^0$ including all its decays is given by an infinite--dimensional vector in Hilbert space:
\begin{eqnarray}
|\tilde{\psi}(t)\rangle &=& a(t) |M^0\rangle+b(t) |\bar M^0\rangle+c(t)|f_1\rangle+d(t)|f_2\rangle+\dots
\end{eqnarray}
where $f_i$ denote all decay products and the state $|\tilde{\psi}(t)\rangle$ is a solution of the Schr\"odinger equation ($\hbar\equiv 1$):
\begin{eqnarray}
\frac{d}{dt}|\tilde{\psi}(t)\rangle&=&-i\hat{H}|\tilde{\psi}(t)\rangle\;
\end{eqnarray}
where $\hat{H}$ is an infinite-dimensional Hamiltonian operator. There is no method
known how to solve this infinite set of coupled differential equations affected
by strong dynamics. The usual procedure is based on restricting to
the time evolution of the components of the flavour eigenstates, $a(t)$ and $b(t)$. Then one uses the Wigner-Weisskopf approximation and can write down an effective Schr\"odinger equation:
\begin{eqnarray}\label{schroedi}
\frac{d}{dt}|\psi(t)\rangle&=&-i\;H |\psi(t)\rangle\;
\end{eqnarray}
where $|\psi\rangle$ is a two dimensional state vector and $H$ is a non-hermitian Hamiltonian. Any non-hermitian Hamiltonian can be written as a sum of two hermitian operators $M,\Gamma$, i.e. $H=M+\frac{i}{2}\Gamma$, where $M$ is the mass-operator, covering the unitary part of the evolution and the operator $\Gamma$ describes the decay property.
The eigenvectors and eigenvalues of the effective Schr\"odinger equation, we denote by:
\begin{eqnarray}
H\;|M_i\rangle &=&\lambda_i\; |M_i\rangle
\end{eqnarray}
with $\lambda_i=m_i+\frac{i}{2} \Gamma_i$ ($c\equiv 1$). For neutral kaons the first solution (with the lower mass) is denoted by $K_S$, the short lived state, and the second eigenvector by $K_L$, the long lived state, as there is a huge difference between the two decay widths $\Gamma_S\simeq 600 \Gamma_L$. For B-mesons the lower mass solution is denoted by $B_L$ with $L$ for light, and the second solution by $B_H$ with $H$ for heavy. For this meson type (as for all the other ones except K-mesons) the decay widths are in good approximation equal, i.e. $\Gamma_L\simeq\Gamma_H$. Thus, the huge difference in two decay  widths is special to K-mesons and this is one reason that they are attractive to various foundational tests, such as e.g. tests for nonlocality \cite{TestableBI,H1,BH1,Genovese2,BGH2,BBGH} or the very working of a quantum eraser \cite{QE1,QE2, HiesmayrComp,HiesmayrKLOE}.

Certainly, the state vector is not normalized for times $t>0$ due to the non-
hermitian part of the dynamics. Different strategies have been developed
to cope with that. In Ref.~\cite{BGH4} the authors developed based on the open quantum formalism a framework that shows that the effect of decay is a kind of decoherence. In this paper we handle only the surviving part of the evolution, but given the framework developed in Ref.~\cite{BGH4} (or similar in Ref.~\cite{Caban}) one can straightforwardly obtain the full quantum information content of the meson systems, e.g. to study Heisenberg's uncertainty in relation with $\mathcal{CP}$ violation in the time evolution~\cite{FoundationPaper,HiesmayrHuber}.

\section*{APPENDIX B: Computation of $I_{kj}^{\left(1\right)}\left(\mathbf{p}_{i};t\right)$}\label{A}

Let us focus our attention on the stochastic average, i.e.:
\begin{eqnarray}
\lefteqn{\mathbb{E}\left[T_{k}^{\left(1\right)*}\left(\mathbf{p}_{f};\mathbf{p}_{i};t\right)T_{j}^{\left(1\right)}\left(\mathbf{p}_{f};\mathbf{p}_{i};t\right)\right]} \qquad \nonumber \\
& & =\mathbb{E}\left[\left(-i\sqrt{\gamma_{m_{k}}}\frac{1}{L^{3}}\int_{0}^{t}dt_{2}\int d\mathbf{x}_{2}w\left(x_{2}\right)e^{-\frac{i}{\hbar}\left[\left(E_{f}^{\left(k\right)}-E_{i}^{\left(k\right)}\right)t_{2}-\left(\mathbf{p}_{f}-\mathbf{p}_{i}\right)\cdot\mathbf{x}_{2}\right]}\right)\right. \nonumber \\
& & \left.\left(i\sqrt{\gamma_{m_{j}}}\frac{1}{L^{3}}\int_{0}^{t}dt_{1}\int d\mathbf{x}_{1}w\left(x_{1}\right)e^{\frac{i}{\hbar}\left[\left(E_{f}^{\left(j\right)}-E_{i}^{\left(j\right)}\right)t_{1}-\left(\mathbf{p}_{f}-\mathbf{p}_{i}\right)\cdot\mathbf{x}_{1}\right]}\right)\right]
\nonumber \\
& & =\sqrt{\gamma_{m_{j}}\gamma_{m_{k}}}\frac{1}{L^{6}}\int_{0}^{t}dt_{1}\int_{0}^{t}dt_{2}\int d\mathbf{x}_{1}\int d\mathbf{x}_{2}\mathbb{E}\left[w\left(x_{1}\right)w\left(x_{2}\right)\right]e^{\frac{i}{\hbar}\left[\left(E_{f}^{\left(j\right)}-E_{i}^{\left(j\right)}\right)t_{1}-\left(\mathbf{p}_{f}-\mathbf{p}_{i}\right)\cdot\mathbf{x}_{1}\right]} \nonumber \\
& & e^{-\frac{i}{\hbar}\left[\left(E_{f}^{\left(k\right)}-E_{i}^{\left(k\right)}\right)t_{2}-\left(\mathbf{p}_{f}-\mathbf{p}_{i}\right)\cdot\mathbf{x}_{2}\right]}.
\end{eqnarray}
The average over the noise is:
\begin{equation} \label{eq:sdtld}
\mathbb{E}\left[w\left(x_{1}\right)w\left(x_{2}\right)\right]=f\left(t_{1}-t_{2}\right)\frac{e^{-\left(\mathbf{x}_{1}-\mathbf{x}_{2}\right)^{2}/4r_{C}^{2}}}{\left(\sqrt{4\pi}r_{C}\right)^{3}},
\end{equation}
where $f\left(t_{1}-t_{2}\right)$ is a generic correlation function characterizing the noise. Therefore we have:
\begin{eqnarray}
\mathbb{E}\left[T_{k}^{\left(1\right)*}\left(\mathbf{p}_{f};\mathbf{p}_{i};t\right)T_{j}^{\left(1\right)}\left(\mathbf{p}_{f};\mathbf{p}_{i};t\right)\right] & = &
\sqrt{\gamma_{m_{j}}\gamma_{m_{k}}}\underset{\equiv C\left(t,\mathbf{p}_{f},\mathbf{p}_{i}\right)}{\underbrace{\int_{0}^{t}\!\!dt_{1}e^{\frac{i}{\hbar}\left(E_{f}^{\left(j\right)}-E_{i}^{\left(j\right)}\right)t_{1}}\int_{0}^{t}\!\!dt_{2}e^{-\frac{i}{\hbar}\left(E_{f}^{\left(k\right)}-E_{i}^{\left(k\right)}\right)t_{2}}f\left(t_{1}-t_{2}\right)}} \nonumber \\
& & \underset{\equiv S\left(\mathbf{p}_{f},\mathbf{p}_{i}\right)}{\underbrace{\frac{1}{L^{6}}\int d\mathbf{x}_{1}\int d\mathbf{x}_{2}\frac{e^{-\left(\mathbf{x}_{1}-\mathbf{x}_{2}\right)^{2}/4r_{C}^{2}}}{\left(\sqrt{4\pi}r_{C}\right)^{3}}e^{-\frac{i}{\hbar}\left(\mathbf{p}_{f}-\mathbf{p}_{i}\right)\cdot\left(\mathbf{x}_{1}-\mathbf{x}_{2}\right)}}}\;.
\end{eqnarray}
We first compute the function $S$:
\begin{equation}
S\left(\mathbf{p}_{f},\mathbf{p}_{i}\right)\equiv\frac{1}{L^{6}}\int_{-\frac{L}{2}}^{+\frac{L}{2}}d\mathbf{x}_{1}\int_{-\frac{L}{2}}^{+\frac{L}{2}}d\mathbf{x}_{2}\frac{e^{-\left(\mathbf{x}_{1}-\mathbf{x}_{2}\right)^{2}/4r_{C}^{2}}}{\left(\sqrt{4\pi}r_{C}\right)^{3}}e^{-\frac{i}{\hbar}\left(\mathbf{p}_{f}-\mathbf{p}_{i}\right)\cdot\left(\mathbf{x}_{1}-\mathbf{x}_{2}\right)}\;.
\end{equation}
Here we wrote explicitly the integration extremes (given by the
box of side $L$). We make the following change of variables:
\begin{equation}
\mathbf{y}=\left(\mathbf{x}_{1}+\mathbf{x}_{2}\right)\;\;\;\textrm{and}\;\;\;\mathbf{x}=\left(\mathbf{x}_{1}-\mathbf{x}_{2}\right).
\end{equation}
The Jacobian of this transformation is $1/2^{3}$ and the integral (in the one dimensional case) changes as follows:
\begin{equation}
\int_{-\frac{L}{2}}^{+\frac{L}{2}}dx_{1}\int_{-\frac{L}{2}}^{+\frac{L}{2}}dx_{2}f\left(x_{1},x_{2}\right)=\frac{1}{2}\int_{0}^{+L}dx\int_{-\left(L-x\right)}^{+\left(L-x\right)}dy\left[f\left(x,y\right)+f\left(-x,y\right)\right].
\end{equation}
In our case we have:
\begin{equation}
f\left(\mathbf{x},\mathbf{y}\right)=\frac{e^{-\mathbf{x}^{2}/4r_{C}^{2}}}{\left(\sqrt{4\pi}r_{C}\right)^{3}}e^{-\frac{i}{\hbar}\left(\mathbf{p}_{f}-\mathbf{p}_{i}\right)\cdot\mathbf{x}}\longrightarrow f\left(\mathbf{x}\right)+f\left(-\mathbf{x}\right)=\frac{e^{-\mathbf{x}^{2}/4r_{C}^{2}}}{\left(\sqrt{4\pi}r_{C}\right)^{3}}2\cos\left[\frac{1}{\hbar}\left(\mathbf{p}_{f}-\mathbf{p}_{i}\right)\cdot\mathbf{x}\right].
\end{equation}
Accordingly, the function $S$ becomes:
\begin{eqnarray}
S\left(\mathbf{p}_{f},\mathbf{p}_{i}\right) & = & \frac{1}{L^{6}}\frac{1}{2^{3}}\int_{0}^{L}d\mathbf{x}\int_{-\left(L-x\right)}^{+\left(L-x\right)}d\mathbf{y}\frac{e^{-\mathbf{x}^{2}/4r_{C}^{2}}}{\left(\sqrt{4\pi}r_{C}\right)^{3}}2\cos\left[\frac{1}{\hbar}\left(\mathbf{p}_{f}-\mathbf{p}_{i}\right)\cdot\mathbf{x}\right] \nonumber \\
& = & \frac{1}{L^{3}}\int_{0}^{L}d\mathbf{x}\frac{e^{-\mathbf{x}^{2}/4r_{C}^{2}}}{\left(\sqrt{4\pi}r_{C}\right)^{3}}2\cos\left[\frac{1}{\hbar}\left(\mathbf{p}_{f}-\mathbf{p}_{i}\right)\cdot\mathbf{x}\right]\frac{1}{2^{3}}\prod_{i=1}^{3}2\frac{\left(L-x_{i}\right)}{L}.
\end{eqnarray}
In the limit $L\rightarrow\infty$ (which we will do later) the term
$\frac{1}{2^{3}}\prod_{i=1}^{3}2\frac{\left(L-x_{i}\right)}{L}\longrightarrow1$, so we can write:
\begin{equation}
S\left(\mathbf{p}_{f},\mathbf{p}_{i}\right)=\frac{1}{L^{3}}\int_{0}^{L}d\mathbf{x}\frac{e^{-\mathbf{x}^{2}/4r_{C}^{2}}}{\left(\sqrt{4\pi}r_{C}\right)^{3}}2\cos\left[\frac{1}{\hbar}\left(\mathbf{p}_{f}-\mathbf{p}_{i}\right)\cdot\mathbf{x}\right]=\frac{1}{L^{3}}\int_{-L}^{+L}d\mathbf{x}\frac{e^{-\mathbf{x}^{2}/4r_{C}^{2}}}{\left(\sqrt{4\pi}r_{C}\right)^{3}}e^{\frac{i}{\hbar}\left(\mathbf{p}_{f}-\mathbf{p}_{i}\right)\cdot\mathbf{x}}.
\end{equation}
We now compute the function $C\left(t,\mathbf{p}_{f},\mathbf{p}_{i}\right)$.
Let us introduce the two variables $a\equiv\frac{1}{\hbar}\left(E_{f}^{\left(j\right)}-E_{i}^{\left(j\right)}\right)$
e $b\equiv-\frac{1}{\hbar}\left(E_{f}^{\left(k\right)}-E_{i}^{\left(k\right)}\right)$ and so we can write
\begin{eqnarray}
C\left(t,\mathbf{p}_{f},\mathbf{p}_{i}\right) & = & \int_{0}^{t}dt_{1}\int_{0}^{t}dt_{2}e^{iat_{1}}e^{ibt_{2}}f\left(t_{1}-t_{2}\right) \nonumber \\
& = & \int_{0}^{t}dt_{1}\int_{0}^{t}dt_{2}e^{\frac{i}{2}\left[\left(a-b\right)\left(t_{1}-t_{2}\right)+\left(a+b\right)\left(t_{1}+t_{2}\right)\right]}f\left(t_{1}-t_{2}\right).
\end{eqnarray}
We now make the change of variables introduced when computing $S$; we define
$u=t_{1}+t_{2}$ and $s=t_{1}-t_{2}$ and focus our attention only
on the case $b=-a$:
\begin{eqnarray} \label{eq:dfghdrty}
C\left(t,\mathbf{p}_{f},\mathbf{p}_{i}\right) & = & \int_{0}^{t}dt_{1}\int_{0}^{t}dt_{2}e^{ia\left(t_{1}-t_{2}\right)}f\left(t_{1}-t_{2}\right) \nonumber \\
& = & \frac{1}{2}\left[\int_{-t}^{0}dse^{ias}f\left(s\right)\int_{-s}^{2t+s}du+\int_{0}^{t}dse^{ias}f\left(s\right)\int_{s}^{2t-s}du\right] \nonumber \\
& = & \int_{-t}^{0}dse^{ias}f\left(s\right)\left(t+s\right)+\int_{0}^{t}dse^{ias}f\left(s\right)\left(t-s\right)=2\int_{0}^{t}ds\cos\left(as\right)f\left(s\right)\left(t-s\right).\qquad
\end{eqnarray}
In the white noise limit $f\left(s\right)=\delta\left(s\right)$ and the function $C$ takes the very simple expression:
\begin{equation}
C_{w}\left(t,\mathbf{p}_{f},\mathbf{p}_{i}\right)=t\;.
\end{equation}
Collecting all pieces together, we have:
\begin{equation}
\mathbb{E}\left[T_{k}^{\left(1\right)*}\left(\mathbf{p}_{f};\mathbf{p}_{i};t\right)T_{j}^{\left(1\right)}\left(\mathbf{p}_{f};\mathbf{p}_{i};t\right)\right]=\sqrt{\gamma_{m_{j}}\gamma_{m_{k}}}C\left(t,\mathbf{p}_{f},\mathbf{p}_{i}\right)\frac{1}{L^{3}}\int_{-L}^{+L}d\mathbf{x}\frac{e^{-\mathbf{x}^{2}/4r_{C}^{2}}}{\left(\sqrt{4\pi}r_{C}\right)^{3}}e^{\frac{i}{\hbar}\left(\mathbf{p}_{f}-\mathbf{p}_{i}\right)\cdot\mathbf{x}}
\end{equation}
and therefore $I_{jk}^{\left(1\right)}\left(\mathbf{p}_{i};t\right)$ becomes:
\begin{equation}
I_{jk}^{\left(1\right)}\left(\mathbf{p}_{i};t\right)=\sqrt{\gamma_{m_{j}}\gamma_{m_{k}}}e^{\frac{i}{\hbar}\left(E_{i}^{\left(j\right)}-E_{i}^{\left(k\right)}\right)t}\sum_{\mathbf{p}_{f}}e^{\frac{i}{\hbar}\left(E_{f}^{\left(k\right)}-E_{f}^{\left(j\right)}\right)t}C\left(t,\mathbf{p}_{f},\mathbf{p}_{i}\right)\frac{1}{L^{3}}\int_{-L}^{+L}d\mathbf{x}\frac{e^{-\mathbf{x}^{2}/4r_{C}^{2}}}{\left(\sqrt{4\pi}r_{C}\right)^{3}}e^{\frac{i}{\hbar}\left(\mathbf{p}_{f}-\mathbf{p}_{i}\right)\cdot\mathbf{x}}.
\end{equation}
We now are in the position to take the limit $L\rightarrow\infty$, so that the above equation becomes:
\begin{eqnarray}
I_{jk}^{\left(1\right)}\left(\mathbf{p}_{i};t\right) & = & \sqrt{\gamma_{m_{j}}\gamma_{m_{k}}}e^{\frac{i}{\hbar}\left(E_{i}^{\left(j\right)}-E_{i}^{\left(k\right)}\right)t}\int d\mathbf{p}_{f}e^{\frac{i}{\hbar}\left(E_{f}^{\left(k\right)}-E_{f}^{\left(j\right)}\right)t}C\left(t,\mathbf{p}_{f},\mathbf{p}_{i}\right) \nonumber \\
& &\frac{1}{\left(2\pi\hbar\right)^{3}}\int_{-\infty}^{+\infty}d\mathbf{x}\frac{e^{-\mathbf{x}^{2}/4r_{C}^{2}}}{\left(\sqrt{4\pi}r_{C}\right)^{3}}e^{\frac{i}{\hbar}\left(\mathbf{p}_{f}-\mathbf{p}_{i}\right)\cdot\mathbf{x}}.
\end{eqnarray}
The last integral can be computed exactly:
\begin{equation}
\int_{-\infty}^{+\infty}d\mathbf{x}\frac{e^{-\mathbf{x}^{2}/4r_{C}^{2}}}{\left(\sqrt{4\pi}r_{C}\right)^{3}}e^{\frac{i}{\hbar}\left(\mathbf{p}_{f}-\mathbf{p}_{i}\right)\cdot\mathbf{x}}=e^{-\frac{\left(\mathbf{p}_{f}-\mathbf{p}_{i}\right)^{2}r_{C}^{2}}{\hbar^{2}}},
\end{equation}
therefore we have:
\begin{equation}
I_{jk}^{\left(1\right)}\left(\mathbf{p}_{i};t\right)=\sqrt{\gamma_{m_{j}}\gamma_{m_{k}}}e^{\frac{i}{\hbar}\left(E_{i}^{\left(j\right)}-E_{i}^{\left(k\right)}\right)t}\int d\mathbf{p}_{f}e^{\frac{i}{\hbar}\left(E_{f}^{\left(k\right)}-E_{f}^{\left(j\right)}\right)t}C\left(t,\mathbf{p}_{f},\mathbf{p}_{i}\right)\frac{1}{\left(2\pi\hbar\right)^{3}}e^{-\frac{\left(\mathbf{p}_{f}-\mathbf{p}_{i}\right)^{2}r_{C}^{2}}{\hbar^{2}}}.
\end{equation}

Now, the Gaussian function has a spread $\sigma = \hbar/r_{C} \simeq 12~\text{eV/c}$, which is very small
 compared to the typical wavelengths entering the oscillatory terms. Indeed, the module of the exponent of that term is:
\begin{equation}
\frac{t}{\hbar}\left|E_{f}^{\left(k\right)}-E_{f}^{\left(j\right)}\right|=\frac{t}{\hbar}\left|\frac{{\bf p}_{f}^{2}}{2m_{k}}-\frac{{\bf p}_{f}^{2}}{2m_{j}}\right|=\frac{t}{2\hbar}\frac{\left|m_{j}-m_{k}\right|}{m_{j}m_{k}}{\bf p}_{f}^{2}\sim2,7\times10^{-16}\left(\textrm{eV/c}\right)^{-2}.
\end{equation}
In the case $j=k$ this is simply zero while for $j\neq k$ we substitute $\left|m_{S}-m_{L}\right|\simeq3.5\cdot10^{-12}\textrm{MeV}/c^{2}$, $m_{L}\simeq m_{S}\simeq498~\textrm{MeV}/c^{2}$ \cite{ParticleDataBook} and, for the kaons at DA$\Phi$NE
%{\bf PLEASE CHECK IF THE NUMBERS WROTE HERE ARE CORRECT,ANTONIO does it make sense to write $m_{L}\simeq m_{S}\simeq500~\textrm{MeV}/c^{2}$???????????}
we have computed $t$ using the maximum possible length $\text{L}= 10~\text{m} $ and taking $v=0,2c$ so that $t=\text{L}/v=1.6\cdot10^{-7}\textrm{s}$. Regarding the term $C\left(t,\mathbf{p}_{f},\mathbf{p}_{i}\right)$ it can be shown, starting by its definition, that it is composed of terms that involve exponential oscillating with a phase of the same order of the one above multiply with the Fourier transform of the noise correlation function $f\left(t\right)$. This means that, like the exponential case we studied just above, even this term changes little the Gaussian width and so we can bring all these terms outside the integral and set $\mathbf{p}_{f}=\mathbf{p}_{i}$. In this way, we need only computing $C\left(t,\mathbf{p}_{i},\mathbf{p}_{i}\right)$, which is the expression as given in Eq.~\eqref{eq:dfghdrty} with $a=-b=0$. So it remains only a Gaussian integral that can be compute explicitly and we get the final result:
\begin{equation}
I_{jk}^{\left(1\right)}\left(\mathbf{p}_{i};t\right)=\sqrt{\gamma_{m_{j}}\gamma_{m_{k}}}\left[2\int_{0}^{t}dsf\left(s\right)\left(t-s\right)\right]\frac{1}{\left(2\pi\right)^{3}}\frac{\pi^{3/2}}{r_{C}^{3}}.
\end{equation}
In the white noise case $f\left(s\right)=\delta\left(s\right)$, which we are mainly interested in, the above formula reduces to:
\begin{equation}
I_{jk}^{\left(1\right)}\left(\mathbf{p}_{i};t\right)=
\sqrt{\gamma_{m_{j}}\gamma_{m_{k}}}\frac{t}{\left(2\pi\right)^{3}}\frac{\pi^{3/2}}{r_{C}^{3}}\;.
\end{equation}
We can notice that this formula is consistent with the relativistic expression first derived in~Ref.~\cite{neutrino}:
\begin{equation}
I_{jk}^{\left(1\right)}\left(\mathbf{p}_{i},s_{i};t\right)=\sqrt{\gamma_{m_{j}}\gamma_{m_{k}}}\frac{m_{j}m_{k}c^{4}}{E_{i}^{\left(j\right)}E_{i}^{\left(k\right)}}\frac{t}{\left(2\pi\right)^{3}}\frac{\pi^{3/2}}{r_{C}^{3}},
\end{equation}
since, in the non relativistic limit: $E_{i}^{\left(j\right)}=\sqrt{\mathbf{p}_{i}^{2}c^{2}+m_{j}^{2}c^{4}}\simeq m_{j}c^{2}$.

\section*{APPENDIX C: Computation of $I_{j}^{\left(2\right)}\left(\mathbf{p}_{i};t\right)$}\label{B}

We now compute
$I_{j}^{\left(2\right)}\left(\mathbf{p}_{i};t\right)\equiv\mathbb{E}\left[T_{j}^{\left(2\right)}\left(\mathbf{p}_{i};\mathbf{p}_{i};t\right)\right]$, with $T_{j}^{\left(2\right)}\left(\mathbf{p}_{i};\mathbf{p}_{i};t\right)$ as given by Eq.~\eqref{eq:sdktbd}.
Using once more relation Eq.~\eqref{eq:sdtld}, we obtain:
\begin{eqnarray}
I_{j}^{\left(2\right)}\left(\mathbf{p}_{i},s_{i};t\right) & = & -\gamma_{m_{j}}\sum_{\mathbf{k}}\underset{=C'\left(t,\mathbf{p}_{i},\mathbf{k}\right)}{\underbrace{\int_{0}^{t}dt_{1}\int_{0}^{t_{1}}dt_{2}e^{\frac{i}{\hbar}\left(E_{i}^{\left(j\right)}-E_{k}^{\left(j\right)}\right)\left(t_{1}-t_{2}\right)}f\left(t_{1}-t_{2}\right)}} \nonumber \\
& & \cdot \underset{=S\left(\mathbf{p}_{i},\mathbf{k}\right)}{\underbrace{\frac{1}{L^{6}}\int d\mathbf{x}_{1}\int d\mathbf{x}_{2}\frac{e^{-\left(\mathbf{x}_{1}-\mathbf{x}_{2}\right)^{2}/4r_{C}^{2}}}{\left(\sqrt{4\pi}r_{C}\right)^{3}}e^{-\frac{i}{\hbar}\left(\mathbf{p}_{i}-\mathbf{k}\right)\cdot\left(\mathbf{x}_{1}-\mathbf{x}_{2}\right)}}}.
\end{eqnarray}
The function $S$ is the same as the one defined and computed in the previous section. The function $C'$ instead is slightly different from the function $C$ previously introduced, the main difference being that in $C$ the second integral goes from
$0$ to $t_{1}$, while for $C'$ it goes from $0$ to $t$. As before it will be interested just in the case $\mathbf{p}_{i}=\mathbf{k}$ and, performing  similar computation, one finds out that:
\begin{equation}
C'\left(t,\mathbf{p}_{i},\mathbf{p}_{i}\right)\equiv\int_{0}^{t}dt_{1}\int_{0}^{t}dt_{2}\theta\left(t_{1}-t_{2}\right)f\left(t_{1}-t_{2}\right)=\int_{0}^{t}dsf\left(s\right)\left(t-s\right).
\end{equation}
Going back to the definition of $I_{j}^{\left(2\right)}\left(\mathbf{p}_{i};t\right)$,
by taking the limit $L\rightarrow\infty$, one obtains:
\begin{equation}
I_{j}^{\left(2\right)}\left(\mathbf{p}_{i};t\right)=-\gamma_{m_{j}}\int d\mathbf{k}C'\left(t,\mathbf{p}_{i},\mathbf{k}\right)\frac{1}{\left(2\pi\hbar\right)^{3}}e^{-\frac{\left(\mathbf{p}_{i}-\mathbf{k}\right)^{2}r_{C}^{2}}{\hbar^{2}}}
\end{equation}
where the Gaussian integral has already been computed in the previous Appendix~B.
At this point, as we did in the previous section, we can make the approximation, inside
the integral, $C'\left(t,\mathbf{p}_{i},\mathbf{k}\right)\simeq C'\left(t,\mathbf{p}_{i},\mathbf{p}_{i}\right)$
and so we get:
\begin{equation}
I_{j}^{\left(2\right)}\left(\mathbf{p}_{i},s_{i};t\right)=-\gamma_{m_{j}}\left[\int_{0}^{t}dsf\left(s\right)\left(t-s\right)\right]\frac{1}{\left(2\pi\right)^{3}}\frac{\pi^{3/2}}{r_{C}^{3}}\;\;\stackrel{\textrm{white noise case}}{\longrightarrow}\;\;-\frac{\gamma_{m_{j}}}{2}\frac{t}{\left(2\pi\right)^{3}}\frac{\pi^{3/2}}{r_{C}^{3}},
\end{equation}
that is still consistent with the relativistic result:
\begin{equation}
I_{j}^{\left(2\right)}\left(\mathbf{p}_{i},s_{i};t\right)=-\frac{1}{2}\gamma_{m_{j}}\frac{m_{j}^{2}c^{4}}{\omega_{p_{i}}^{2\left(j\right)}}\frac{t}{\left(2\pi\right)^{3}}\frac{\pi^{3/2}}{r_{C}^{3}}\;.
\end{equation}

\section*{APPENDIX D: A computation with wave packets}

Let us start from the formula Eq.~\eqref{eq:Pdouble}:
\begin{eqnarray}
P\left(F_{l};F_{r}\right) & \equiv & \sum_{\mathbf{p}_{l},\mathbf{p}_{r}}\mathbb{E}\left|A\left(F_{l},\mathbf{p}_{l};F_{r},\mathbf{p}_{r}\right)\right|^{2}=\sum_{j,k,j',k'=S,L}\alpha_{jk}\beta_{j}^{*}\gamma_{k}^{*}\alpha_{j'k'}^{*}\beta_{j'}\gamma_{k'}\times \nonumber \\
& \times & \mathbb{E} \left\{\left[\sum_{\mathbf{p}_{l}}\left\langle K_{j'},\mathbf{p}_{l}\left|U_{j'}\left(t_{l}\right)\right|K_{j'},-\mathbf{p}_{i}\right\rangle ^{*}\left\langle K_{j},\mathbf{p}_{l}\left|U_{j}\left(t_{l}\right)\right|K_{j},-\mathbf{p}_{i}\right\rangle \right]\right.\times\nonumber \\
& \times & \left.\left[\sum_{\mathbf{p}_{r}}\left\langle K_{k'},\mathbf{p}_{r}\left|U_{k'}\left(t_{r}\right)\right|K_{k'},\mathbf{p}_{i}\right\rangle ^{*}\left\langle K_{k},\mathbf{p}_{r}\left|U_{k}\left(t_{r}\right)\right|K_{k},\mathbf{p}_{i}\right\rangle \right]\right\}\;.
\end{eqnarray}
Here we want to make clear why we can neglect all the terms involving the correlation between terms in the first square bracket with terms in the second square bracket. We recall that:
\begin{equation}
\left\langle K_{j},\mathbf{p}_{f}\left|U_{j}\left(t\right)\right|K_{j},\mathbf{p}_{i}\right\rangle = e^{-\frac{i}{\hbar}E_{f}^{\left(j\right)}t}\left[\delta_{\mathbf{p}_{f},\mathbf{p}_{i}}+\underset{\textrm{1 noise }}{\underbrace{T_{j}^{\left(1\right)}\left(\mathbf{p}_{f};\mathbf{p}_{i};t\right)}}+\underset{\textrm{2 noise }}{\underbrace{T_{j}^{\left(2\right)}\left(\mathbf{p}_{f};\mathbf{p}_{i};t\right)}}\right]\;.
\end{equation}
If we substitute this in the second and third lines and we take only the contribution up to the second order we get:
\[
\mathbb{E} \left\{\left[\sum_{\mathbf{p}_{l}}\left\langle K_{j'},\mathbf{p}_{l}\left|U_{j'}\left(t_{l}\right)\right|K_{j'},-\mathbf{p}_{i}\right\rangle ^{*}\left\langle K_{j},\mathbf{p}_{l}\left|U_{j}\left(t_{l}\right)\right|K_{j},-\mathbf{p}_{i}\right\rangle \right]\right.\times
\]
\[
\times\left.\left[\sum_{\mathbf{p}_{r}}\left\langle K_{k'},\mathbf{p}_{r}\left|U_{k'}\left(t_{r}\right)\right|K_{k'},\mathbf{p}_{i}\right\rangle ^{*}\left\langle K_{k},\mathbf{p}_{r}\left|U_{k}\left(t_{r}\right)\right|K_{k},\mathbf{p}_{i}\right\rangle \right]\right\}=
\]
\[
=e^{\frac{i}{\hbar}\left(E_{i}^{\left(j'\right)}-E_{i}^{\left(j\right)}\right)t_{l}}e^{\frac{i}{\hbar}\left(E_{i}^{\left(k'\right)}-E_{i}^{\left(k\right)}\right)t_{r}}\left\{ 1+\mathbb{E}\left[T_{j'}^{\left(2\right)*}\left(-\mathbf{p}_{i};-\mathbf{p}_{i};t_{l}\right)\right]+\mathbb{E}\left[T_{j}^{\left(2\right)}\left(-\mathbf{p}_{i};-\mathbf{p}_{i};t_{l}\right)\right]+\right.\]
\[
+\mathbb{E}\left[T_{k'}^{*\left(2\right)}\left(\mathbf{p}_{i};\mathbf{p}_{i};t_{r}\right)\right]+\mathbb{E}\left[T_{k}^{\left(2\right)}\left(\mathbf{p}_{i};\mathbf{p}_{i};t_{r}\right)\right]+\]
\[
+\mathbb{E}\left[T_{j'}^{\left(1\right)*}\left(-\mathbf{p}_{i};-\mathbf{p}_{i};t_{l}\right)T_{k'}^{\left(1\right)*}\left(\mathbf{p}_{i};\mathbf{p}_{i};t_{r}\right)\right]+\mathbb{E}\left[T_{j'}^{\left(1\right)*}\left(-\mathbf{p}_{i};-\mathbf{p}_{i};t_{l}\right)T_{k}^{\left(1\right)}\left(\mathbf{p}_{i};\mathbf{p}_{i};t_{r}\right)\right]+\]
\[
\left.+\mathbb{E}\left[T_{j}^{\left(1\right)}\left(-\mathbf{p}_{i};-\mathbf{p}_{i};t_{l}\right)T_{k'}^{\left(1\right)*}\left(\mathbf{p}_{i};\mathbf{p}_{i};t_{r}\right)\right]+\mathbb{E}\left[T_{j}^{\left(1\right)}\left(-\mathbf{p}_{i};-\mathbf{p}_{i};t_{l}\right)T_{k}^{\left(1\right)}\left(\mathbf{p}_{i};\mathbf{p}_{i};t_{r}\right)\right]\right\} +\]
\[
+e^{\frac{i}{\hbar}\left(E_{i}^{\left(k'\right)}-E_{i}^{\left(k\right)}\right)t_{r}}\sum_{\mathbf{p}_{l}}e^{\frac{i}{\hbar}\left(E_{l}^{\left(j'\right)}-E_{l}^{\left(j\right)}\right)t_{l}}\mathbb{E}\left[T_{j'}^{\left(1\right)*}\left(\mathbf{p}_{l};-\mathbf{p}_{i};t_{l}\right)T_{j}^{\left(1\right)}\left(\mathbf{p}_{l};-\mathbf{p}_{i};t_{l}\right)\right]+\]
\begin{equation}\label{eq:LONG}
+e^{\frac{i}{\hbar}\left(E_{i}^{\left(j'\right)}-E_{i}^{\left(j\right)}\right)t_{l}}\sum_{\mathbf{p}_{r}}e^{\frac{i}{\hbar}\left(E_{r}^{\left(k'\right)}-E_{r}^{\left(k\right)}\right)t_{r}}\mathbb{E}\left[T_{k'}^{\left(1\right)*}\left(\mathbf{p}_{r};\mathbf{p}_{i};t_{r}\right)T_{k}^{\left(1\right)}\left(\mathbf{p}_{r};\mathbf{p}_{i};t_{r}\right)\right].
\end{equation}
We can see that the first and the last two lines contain pieces that involve
only terms referring to the same particle (left or right). But there are also pieces containing
the product of terms which refer to one particle and terms which refer to the other particle. This pieces
seems unphysical, especially because they are present
even if we take a separable state as the initial state. And the fact that the correlation
between left terms and
right terms could play an important role even for separable
states sounds suspicious. Indeed this problem arises because we
had oversimplified our analysis: by using plane waves as states of the
particles, we are working with totally delocalized states, which is the origin of the correlation between terms of the left and the right
particle. Anyway this is not what actually happens in
laboratories: in such cases, the state of the system is described
by wave packets well localized in space and we will show now that,
with such an assumption, the suspicious pieces cancel. To do that, we have to
replace this plane waves with wave packets. Therefore, in place of
the two initial states $\left|\mathbf{p}_{i}\right\rangle $ and $\left|-\mathbf{p}_{i}\right\rangle$,
we now consider:
\begin{equation}
\left|f\right\rangle =\sum_{\mathbf{p}_{i}}f\left(\mathbf{p}_{i}\right)\left|\mathbf{p}_{i}\right\rangle \;\;\;\textrm{and}\;\;\;\left|g\right\rangle =\sum_{\mathbf{p}_{i}}g\left(\mathbf{p}_{i}\right)\left|\mathbf{p}_{i}\right\rangle,
\end{equation}
where $\left|f\right\rangle $ ($\left|g\right\rangle $)
it is a wave packet propagating in the left (right) direction. Regarding the final states, we will continue to take them as plane waves, because, as before, we will sum over them. Taken different initial states implies that the matrix element $\left\langle K_{j},\mathbf{p}_{f}\left|U_{j}\left(t\right)\right|K_{j},\mathbf{p}_{i}\right\rangle$ become:
\begin{equation}
\left\langle K_{j},\mathbf{p}_{f}\left|U_{j}\left(t\right)\right|K_{j},f\right\rangle =e^{-\frac{i}{\hbar}E_{f}^{\left(j\right)}t}\sum_{\mathbf{p}_{i}}f\left(\mathbf{p}_{i}\right)\left[\delta_{\mathbf{p}_{f},\mathbf{p}_{i}}+\underset{\textrm{1 noise }}{\underbrace{T_{j}^{\left(1\right)}\left(\mathbf{p}_{f};\mathbf{p}_{i};t\right)}}+\underset{\textrm{2 noise }}{\underbrace{T_{j}^{\left(2\right)}\left(\mathbf{p}_{f};\mathbf{p}_{i};t\right)}}\right]
\end{equation}
and the same happen with the matrix element containing $\left|g\right\rangle $. Let's see how change the bad terms, for example the one that before was:
\begin{equation}
e^{\frac{i}{\hbar}\left(E_{i}^{\left(j'\right)}-E_{i}^{\left(j\right)}\right)t_{l}}e^{\frac{i}{\hbar}\left(E_{i}^{\left(k'\right)}-E_{i}^{\left(k\right)}\right)t_{r}}\mathbb{E}\left[T_{j}^{\left(1\right)}\left(-\mathbf{p}_{i};-\mathbf{p}_{i};t_{l}\right)T_{k'}^{\left(1\right)*}\left(\mathbf{p}_{i};\mathbf{p}_{i};t_{r}\right)\right].
\end{equation}
Looking at the formula for the correlation Eq.~\eqref{eq:LONG}, we know that this term
came out taking the delta (i.e. the zero order contribution) for the
bracket labeled by $j'$ and $k$ and taking the first order term $T^{\left(1\right)}$
for the ones labeled by the index $j$ and $k'$. If we take now the corresponding terms
we get $f^{*}\left(\mathbf{p}_{l}\right)$, $g\left(\mathbf{p}_{r}\right)$ instead of $\delta_{\mathbf{p}_{l},\mathbf{p}_{i}}$, $\delta_{\mathbf{p}_{r},\mathbf{p}_{i}}$ and $\sum_{\mathbf{p}_{1}}f\left(\mathbf{p}_{1}\right)T_{j}^{\left(1\right)}\left(\mathbf{p}_{l};\mathbf{p}_{1};t_{l}\right)$, $\sum_{\mathbf{p}_{2}}g^{*}\left(\mathbf{p}_{2}\right)T_{k'}^{\left(1\right)*}\left(\mathbf{p}_{r};\mathbf{p}_{2};t_{r}\right)$ instead of $T_{j}^{\left(1\right)}\left(\mathbf{p}_{i};\mathbf{p}_{1};t_{l}\right)$, $T_{k'}^{\left(1\right)*}\left(\mathbf{p}_{i};\mathbf{p}_{2};t_{r}\right)$ and so it becomes:
\begin{equation}
\sum_{\mathbf{p}_{l},\mathbf{p}_{r}}e^{\frac{i}{\hbar}\left(E_{l}^{\left(j'\right)}-E_{l}^{\left(j\right)}\right)t_{l}}e^{\frac{i}{\hbar}\left(E_{r}^{\left(k'\right)}-E_{r}^{\left(k\right)}\right)t_{r}}f^{*}\left(\mathbf{p}_{l}\right)g\left(\mathbf{p}_{r}\right)\sum_{\mathbf{p}_{1},\mathbf{p}_{2}}f\left(\mathbf{p}_{1}\right)g^{*}\left(\mathbf{p}_{2}\right)\mathbb{E}\left[T_{j}^{\left(1\right)}\left(\mathbf{p}_{l};\mathbf{p}_{1};t_{l}\right)T_{k'}^{\left(1\right)*}\left(\mathbf{p}_{r};\mathbf{p}_{2};t_{r}\right)\right]\;.
\end{equation}
Regarding this expression the interesting piece is:
\begin{equation}
Z\equiv\sum_{\mathbf{p}_{1},\mathbf{p}_{2}}f\left(\mathbf{p}_{1}\right)g^{*}\left(\mathbf{p}_{2}\right)\mathbb{E}\left[T_{j}^{\left(1\right)}\left(\mathbf{p}_{l};\mathbf{p}_{1};t_{l}\right)T_{k'}^{\left(1\right)*}\left(\mathbf{p}_{r};\mathbf{p}_{2};t_{r}\right)\right].
\end{equation}
Using Eq.~\eqref{eq:T1}:
\begin{eqnarray}
Z & = & \sqrt{\gamma_{m_{j}}\gamma_{m_{k'}}}\, \mathbb{E}\left[\int_{0}^{t_{l}}dt_{1}\int d\mathbf{x}_{1}w\left(x_{1}\right)\left\langle \Omega\right|b_{j\mathbf{p}_{l}}\psi_{jI}^{\dagger}\left(x_{1}\right)\psi_{jI}\left(x_{1}\right)\left|f\right\rangle\right.\times \nonumber \\
& \times & \left.\int_{0}^{t_{r}}dt_{2}\int d\mathbf{x}_{2}w\left(x_{2}\right)\left\langle \Omega\right|b_{k'\mathbf{p}_{r}}\psi_{k'I}^{\dagger}\left(x_{2}\right)\psi_{k'I}\left(x_{2}\right)\left|g\right\rangle \right]\;.
\end{eqnarray}
We can focus on compute the matrix elements. For example the second
one is:
\begin{equation}
\left\langle \Omega\right|b_{k'\mathbf{p}_{r}}\psi_{k'I}^{\dagger}\left(x_{2}\right)\psi_{k'I}\left(x_{2}\right)\left|g\right\rangle =\left\langle \Omega\right|b_{k'\mathbf{p}_{r}}\psi_{k'I}^{\dagger}\left(x_{2}\right)e^{\frac{i}{\hbar}H_{S}t_{2}}\psi_{k'}\left(\mathbf{x}_{2}\right)e^{-\frac{i}{\hbar}H_{S}t_{2}}\left|g\right\rangle.
\end{equation}
Now we can introduce $\left|g,t_{2}\right\rangle \equiv e^{-\frac{i}{\hbar}H_{S}t_{2}}\left|g\right\rangle $, that is just the wave packet evaluated up to the time $t_{2}$ with free evolution. From the experiment we know that the kaons are always found one at left one at right. This means that the wavepackets associated to these kaons does not spread too much compare to their distance from the source, i.e. $g\left(\mathbf{x},t_{2}\right)=\left\langle\mathbf{x}\right.\left|g,t_{2}\right\rangle$.\footnote{One can claim that the wavepacket observed in the laboratory is the one given by the complete evolution, taken into account also the noise, and so it is different from $g\left(\mathbf{x},t\right)$ that is the wavepacket evolved with the free Hamiltonian. Anyway it is well know from the literature that the effect of the noise on the spread of the wavefunction in typical times of this experiment is negligible, so we can be sure that if the real wavepacket remains confined, the same happens also for $g\left(\mathbf{x},t\right)$.}

Now we can use the key relation:
\begin{equation}
\left|g,t_{2}\right\rangle =\int d\mathbf{x}g\left(\mathbf{x},t_{2}\right)\psi_{k'}^{\dagger}\left(\mathbf{x}\right)\left|\Omega\right\rangle
\end{equation}
and so:

\[
\left\langle \Omega\right|b_{k'\mathbf{p}_{r}}\psi_{k'I}^{\dagger}\left(x_{2}\right)\psi_{k'I}\left(x_{2}\right)\left|g\right\rangle =\int d\mathbf{x}g\left(\mathbf{x},t_{2}\right)\left\langle \Omega\right|b_{k'\mathbf{p}_{r}}\psi_{k'I}^{\dagger}\left(x_{2}\right)e^{\frac{i}{\hbar}H_{S}t_{2}}\psi_{k'}\left(\mathbf{x}_{2}\right)\psi_{k'}^{\dagger}\left(\mathbf{x}\right)\left|\Omega\right\rangle =
\]
\begin{equation}
=g\left(\mathbf{x}_{2},t_{2}\right)\left\langle \Omega\right|b_{k'\mathbf{p}_{r}}\psi_{k'I}^{\dagger}\left(x_{2}\right)e^{\frac{i}{\hbar}H_{S}t_{2}}\left|\Omega\right\rangle =g\left(\mathbf{x}_{2},t_{2}\right)\frac{e^{\frac{i}{\hbar}\left(E_{r}^{\left(k'\right)}t_{2}-\mathbf{p}_{r}\cdot\mathbf{x}_{2}\right)}}{\sqrt{L^{3}}}.
\end{equation}
Doing the same for the other matrix element and inserting everything in $Z$ we get:
\begin{eqnarray}
Z & = & \sqrt{\gamma_{m_{j}}\gamma_{m_{k'}}}\int_{0}^{t_{l}}dt_{1}\int_{0}^{t_{r}}dt_{2}\int d\mathbf{x}_{1}\int d\mathbf{x}_{2}\frac{e^{\frac{i}{\hbar}\left(E_{l}^{\left(j\right)}t_{1}-\mathbf{p}_{l}\cdot\mathbf{x}_{1}\right)}e^{\frac{i}{\hbar}\left(E_{r}^{\left(k'\right)}t_{2}-\mathbf{p}_{r}\cdot\mathbf{x}_{2}\right)}}{L^{3}}\times \nonumber \\
& \times & f\left(\mathbf{x}_{1},t_{1}\right)g\left(\mathbf{x}_{2},t_{2}\right)\mathbb{E}\left[w\left(x_{1}\right)w\left(x_{2}\right)\right]
\end{eqnarray}

and using the correlation (in the white noise case) $\mathbb{E}\left[w\left(x_{1}\right)w\left(x_{2}\right)\right]=\delta\left(t_{1}-t_{2}\right)\frac{e^{-\left(\mathbf{x}_{1}-\mathbf{x}_{2}\right)^{2}/4r_{C}^{2}}}{\left(\sqrt{4\pi}r_{C}\right)^{3}}$:

\begin{eqnarray}
Z & = & \sqrt{\gamma_{m_{j}}\gamma_{m_{k'}}}\frac{1}{L^{3}}\int_{0}^{\min\left(t_{l},t_{r}\right)}dt_{1}e^{\frac{i}{\hbar}\left(E_{l}^{\left(j\right)}+E_{r}^{\left(k'\right)}\right)t_{1}}\times \nonumber \\
& \times & \int d\mathbf{x}_{1}\int d\mathbf{x}_{2}e^{-\frac{i}{\hbar}\mathbf{p}_{l}\cdot\mathbf{x}_{1}}e^{-\frac{i}{\hbar}\mathbf{p}_{r}\cdot\mathbf{x}_{2}}f\left(\mathbf{x}_{1},t_{1}\right)g\left(\mathbf{x}_{2},t_{1}\right)\frac{e^{-\left(\mathbf{x}_{1}-\mathbf{x}_{2}\right)^{2}/4r_{C}^{2}}}{\left(\sqrt{4\pi}r_{C}\right)^{3}}\;.
\end{eqnarray}

We recall that $r_{C}=10^{-7}\textrm{m}$ so, a part for the starting instants,
the two wave packet are always far away and the integral is small.
This doesn't happen for the terms where the correlation is taken between
noise related to the same particle, because in such case we end up
with terms containing pieces like $f\left(\mathbf{x}_{1},t_{1}\right)f\left(\mathbf{x}_{2},t_{1}\right)$
or $g\left(\mathbf{x}_{1},t_{1}\right)g\left(\mathbf{x}_{2},t_{1}\right)$
inside the integrals, that gives an important contribution even for large time.

\end{document}